\pdfoutput=1

\documentclass[11pt]{article}

\usepackage[preprint]{acl}
\usepackage{hyperref}
\usepackage{url}
\usepackage{algorithm}
\usepackage{algorithmic}
\usepackage{amsmath,amssymb}
\usepackage{tabularx}
\usepackage{pifont}
\usepackage{amsmath,amsthm,amssymb,amsfonts}

\usepackage{newfloat}
\usepackage{listings}
\usepackage{times}  
\usepackage{helvet}  
\usepackage{times}
\usepackage{latexsym}
\usepackage{times}
\usepackage{latexsym}
\usepackage{hyperref}
\definecolor{ccr}{RGB}{30,70,255}  
\usepackage{hyperref}
\hypersetup{hypertex=true,
    colorlinks=true,
    linkcolor=ccr,
    anchorcolor=ccr,
    citecolor=ccr}
\usepackage{times}
\usepackage{latexsym}
\usepackage[utf8]{inputenc}
\usepackage{array}
\usepackage{multirow,booktabs}
\usepackage{booktabs}
\usepackage{microtype}

\usepackage{inconsolata}
\usepackage{CJKutf8}
\usepackage{amsmath}
\usepackage{xspace}
\usepackage{graphicx}
\usepackage{algorithm}
\usepackage{algorithmic}
\usepackage{footnote}
\usepackage{makecell}
\usepackage{makecell}
\usepackage{booktabs}
\usepackage{multirow}
\usepackage{utfsym}
\usepackage{enumitem}
\usepackage{subfigure}
\usepackage{amsmath,amssymb}
\usepackage{enumitem}
\usepackage{balance}
\usepackage{bm}
\usepackage{array}

\usepackage{booktabs}
\usepackage[T1]{fontenc}

\usepackage[utf8]{inputenc}

\usepackage{microtype}

\usepackage{inconsolata}

\usepackage{graphicx}

\title{RuleRAG: Rule-Guided Retrieval-Augmented Generation 
with Language Models 
for Question Answering}

\author{Zhongwu Chen$^{1}$, Chengjin Xu$^{2*}$, Dingmin Wang$^{3*}$, Zhen Huang$^{1*}$, Yong Dou$^{1}$, Xuhui Jiang$^{2}$, Jian Guo$^{2}$\\
$^{1}$National Key Laboratory of Parallel and Distributed Computing, 
\\\textcolor{white}{$^{1}$}College of Computer Science and Technology,
National University of Defense Technology
\\$^{2}$IDEA Research, International Digital Economy Academy \\$^{3}$Department of Computer Science, University of Oxford\\
\texttt{\{chenzhongwu20, huangzhen, yongdou\}@nudt.edu.cn},
\\
\texttt{\{xuchengjin, jiangxuhui, guojian\}@idea.edu.cn, \{dingmin.wang\}@cs.ox.ac.uk} \\
}

\begin{document}
\maketitle
\begin{abstract}
Retrieval-augmented generation (RAG)
has shown promising potential in knowledge intensive question answering (QA).
\emph{However, existing approaches only consider the query itself, neither specifying the retrieval preferences for the retrievers nor informing the generators of how to refer to the retrieved documents for the answers, which poses a significant challenge to the QA performance.}
To address these issues, we propose Rule-guided Retrieval-Augmented Generation with LMs, which explicitly introduces rules for in-context learning (RuleRAG-ICL) to guide retrievers to recall related documents in the directions of rules and uniformly guide generators to reason attributed by the same rules.
The combination of queries and rules can be used as fine-tuning data to update retrievers and generators, achieving better rule-based instruction-following ability (RuleRAG-FT).
\emph{Moreover, most existing RAG datasets were constructed without considering rules and Knowledge Graphs (KGs) are recognized as providing high-quality rules.}
Therefore, we construct five rule-aware RAG benchmarks for QA, RuleQA, based on KGs to stress the significance of retrieval and reasoning with rules.
Experiments on RuleQA demonstrate RuleRAG-ICL improves the retrieval quality of +89.2\% in Recall@10 and answer accuracy of +103.1\% in Exact Match, and RuleRAG-FT yields more enhancement.
In addition, experiments on four existing RAG datasets show 
RuleRAG is also effective by offering rules in RuleQA to them, further proving the generalization of rule guidance in RuleRAG. 
Code and RuleQA are at \url{https://anonymous.4open.science/r/RuleRAG}.
\end{abstract}

\section{Introduction}

Large language models (LLMs) have achieved the impressive capability of language generation and knowledge learning~\citep{NEURIPS2020_1457c0d6,NEURIPS2022_b1efde53}.
Despite the success, the full-parametric knowledge in LLMs 
struggles to precisely manipulate fine-grained queries, especially in knowledge-intensive tasks~\citep{2023-active,shao-etal-2023-enhancing}.
As complementary, 
RAG
shows superior performance in many NLP tasks, such as
open-domain 
QA~\citep{
trivedi-etal-2023-interleaving} and natural language inference~\citep{qin-etal-2023-chatgpt}.\looseness=-1

\begin{figure*}[h]
    \centering
    \includegraphics
    [width=1\linewidth]
    {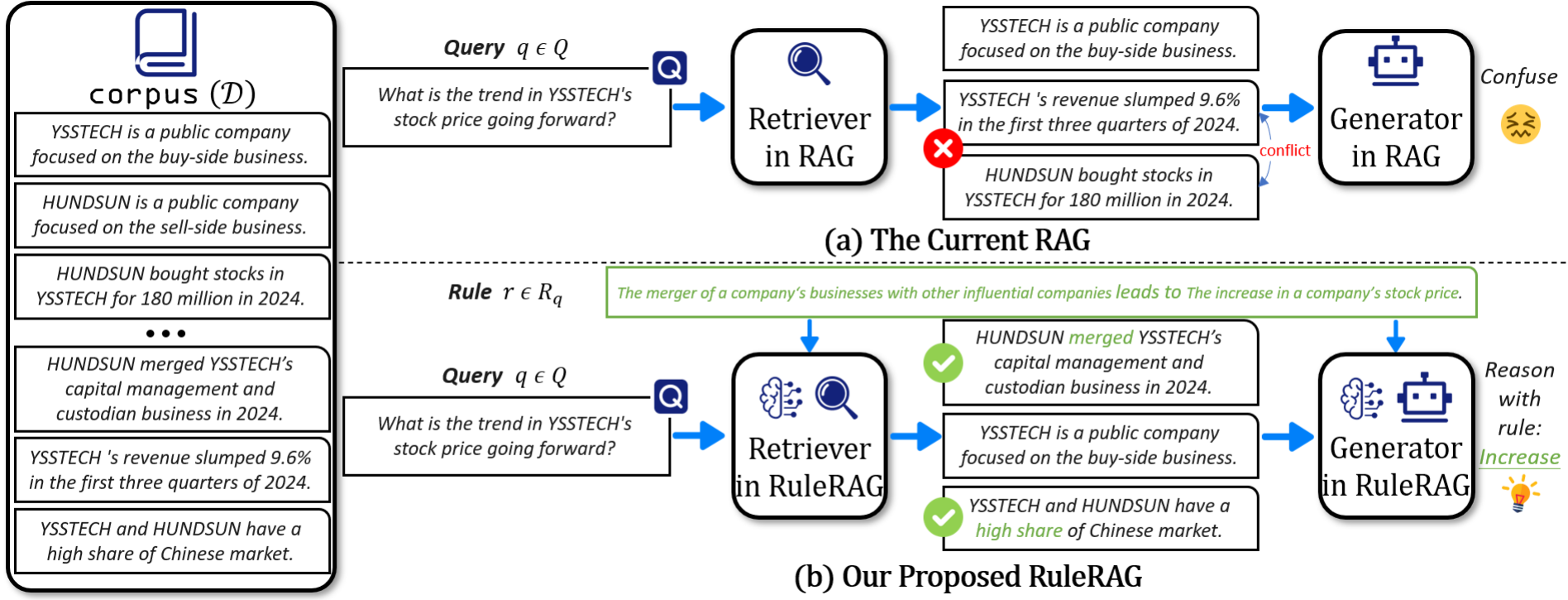}
    \caption{(a) Without the help of rules, the current RAG can only retrieve documents about some keywords, rather than the overall semantics of the query, and thus get confused in answering. (b) Guided by the attributable rule $r$, our proposed RuleRAG retrieves logically supportive documents and then reasons the correct answer .\looseness=-1
    }
    \vspace{-0.5cm}
    \label{f1}
\end{figure*}

However, two high-level issues exist in the current RAG. 
First, in the retrieval phase, the retrievers rely on word-level matching, and thus can not guarantee that the recalled information is always pertinent to the query answering. The reason is many retrievers are trained on unsupervised text
or trained end-to-end,
leading to their insufficiency in retrieving the necessary statements for reasoning~\citep{behnamghader-etal-2023-retriever}. 
Secondly, in the generation phase, the LLMs in the current RAG are not specifically informed of how to exploit noisy retrieved content properly, since relationships between a wide range of facts are rarely explicitly ``pointed out'' and ``supervised'' in the pre-training corpora of LLMs.
Even if answered correctly, they still lead to implicit attribution processes that are difficult to explain and verify. Therefore, the current RAG is neither inherently trained to retrieve 
along reasonable retrieval directions nor organically attribute retrieved content to answers.


While answering knowledge-intensive queries, 
a priori rules instead of simply matching words can genuinely 
capture internal logical patterns among complicated knowledge.
Some works incorporate rules into LLMs to handle the addition of numbers~\citep{hu2024casebased} or industrial tasks~\citep{zhang2024ruaglearnedruleaugmentedgenerationlarge}, but there is currently no exploration of introducing rules into RAG for QA.
As shown in Figure~\ref{f1}, the query is 
\emph{What is the trend in YSSTECH's stock price going forward?}.
Current retrievers recall many documents that contribute nothing to answering because of blind retrieval.
By contrast, financial KGs provide a rule that 
\emph{The merger of a company's businesses with other influential companies} leads to \emph{The increase in a company's stock price}.
Therefore, we can leverage this rule to conduct more targeted retrieval and offer documents that can better support question answering.
\looseness=-1


Upon the above motivation, we propose RuleRAG, Rule-guided Retrieval-Augmented Generation, which enables to both retrieve documents and reason answers with the guidance of rules.
Compared to standard RAG, which relies on finding the precise statement of the query to be answered, training-free RuleRAG-ICL requires the introduction of 
rules in the input sides of the retrievers to retrieve and generators to reason. 
To boost the rule-following reasoning ability, 
we further design rule-guided fine-tuning (RGFT) to retrofit retrievers and generators (RuleRAG-FT).\looseness=-1
However, although rules are common and valuable for the QA task, most existing RAG datasets were constructed without considering rules.
KGs are widely known to provide question-answer pairs and high-coverage rules, so we newly construct five rule-aware QA benchmarks (RuleQA) from five KGs to offer rich data.
The strong performance of KG rule mining algorithms also ensures the acquisition of rules with high confidence.
In addition, the answers in RuleQA
require reasoning based on rules rather than directly repeating retrieved facts like existing RAG datasets,
so RuleQA is knowledge-intensive and more challenging.
Experiments on RuleQA show that, under several retrieval and generation configurations, RuleRAG-ICL offers considerable performance gains with the individual guidance of rules by in-context learning and RuleRAG-FT achieves further improvements by fine-tuning retrievers and generators. RuleRAG-FT can be extrapolated to unseen rules without retraining, confirming the superiority of rules to retrieve and reason in RAG.
We also introduce RuleRAG into an advanced model CoK to emphasize the paradigm of Rule-guided RAG is suitable for different RAG methods.
Moreover, we conduct experiments on four existing RAG datasets.
As a result, RuleRAG is beneficial for our constructed RuleQA and for introducing rules to existing RAG datasets.
RuleRAG-CoK shows the attribution of the advanced RAG-based variant of RuleRAG under RuleQA and existing RAG methods.




\section{Related Work}
\subsection{Retrieval-Augmented Generation}
In RAG, the retrieval module explicitly augments the generation module with external knowledge banks~\citep{REALM}. 
Retrieval approaches include sparse retrievers based on sparse bag-of-words, dense retrievers based on dense vectors and more complex hybrid search algorithms~\citep{li-etal-2023-citadel}.
RAG is widely adopted to complement the LLM parametric knowledge
along different stages~\citep{gao2024retrievalaugmentedgenerationlargelanguage}, including pre-training stage (RETRO~\citep{pmlr-v162-borgeaud22a}, COG~\citep{lan2023copy}, Atlas~\citep{Atlas}), fine-tuning stage (SURGE~\citep{kang2023knowledgegraphaugmentedlanguagemodels}, Self-RAG~\citep{asai2023selfrag}, CoN~\citep{yu2023chainofnoteenhancingrobustnessretrievalaugmented}) and inference stage (KnowledGPT~\citep{wang2023knowledgptenhancinglargelanguage}, DSP~\citep{khattab2023demonstratesearchpredictcomposingretrievallanguage}, RoG~\citep{luo2024reasoning}). 
\looseness=-1

 \begin{figure*}[!]
    \centering
    \includegraphics
    [width=1\linewidth]
    {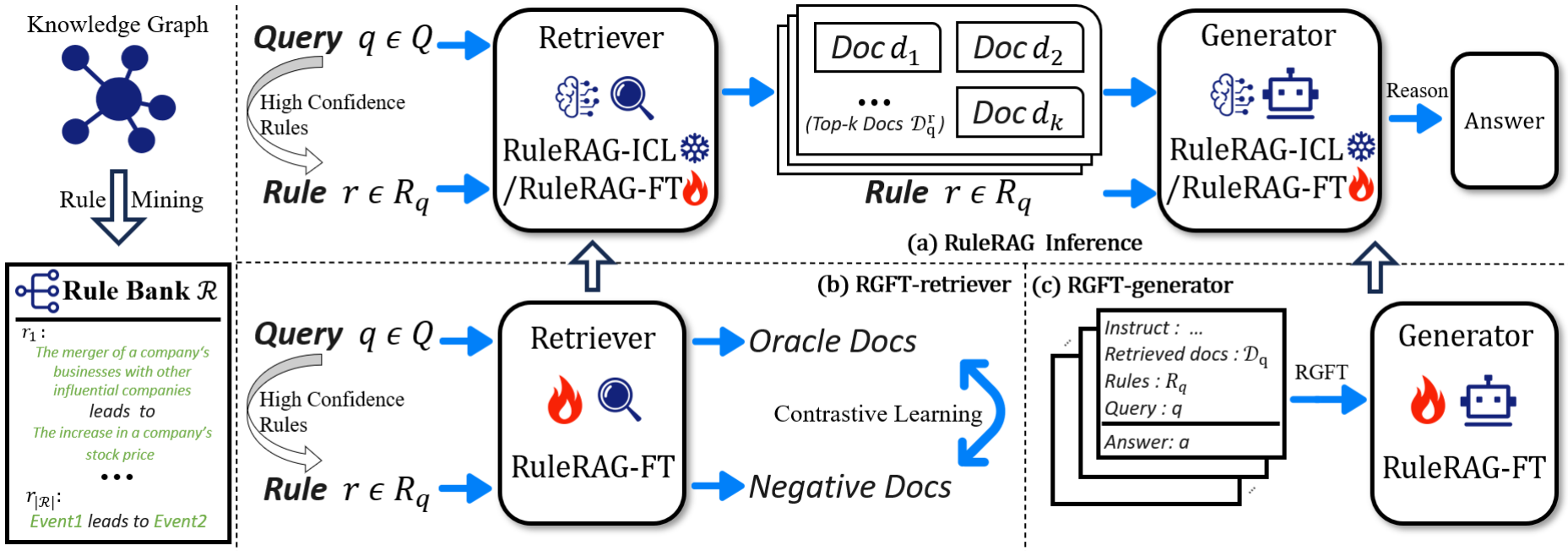}
    \caption{ The framework of our proposed RuleRAG. RuleRAG-ICL relies on in-context learning with the guidance of rules. RuleRAG-FT involves fine-tuning retrievers and generators ahead. (a) The unified RuleRAG inference process. (b) Rule-guided retriever fine-tuning (RGFT-retriever). (c) Rule-guided generator fine-tuning (RGFT-generator).\looseness=-1}
    
    \vspace{-0.3cm}
    \label{mainfigure}
\end{figure*}

\subsection{Knowledge-intensive QA}
In the realm of QA, queries are viewed as knowledge-intensive if we need access to external corpora~\citep{thorne-etal-2018-fever}.
Assuming that documents in the corpora include the exact answers, RAFT~\citep{RAFT} and RA-DIT~\citep{lin2024radit} fine-tune LLMs by concatenating documents and queries as prompts.
However, many answers to factual queries are hidden in \emph{semantically dissimilar but logically related documents}, which need to be retrieved and reasoned with the guidance of rules. Our constructed RuleQA simulates this circumstance, while most existing RAG datasets lack rules.
Recently, ~\citet{wu2024easilyirrelevantinputsskew} investigates mitigating misleading irrelevant interference;
~\citet{sun2024instructionfollowingevaluatinginferential} 
only discusses the rule-following abilities of LLMs without retrieval and ignores how to obtain rules. In contrast, our proposed RuleRAG involves a more comprehensive consideration of mining rules, retrieving documents and reasoning answers.
\looseness=-1


%




\section{Proposed Method: RuleRAG}

\subsection{RuleRAG-ICL}
Figure~\ref{mainfigure} (a) illustrates the inference flow of RuleRAG-ICL.
Given a query $q$ $\in$ $\rm{Q}$, 
we first leverage Sentence-BERT~\citep{reimers-gurevych-2019-sentence} to capture the semantic similarity between $q$ and candidate rules. The highest N rules among those whose scores exceed a certain threshold $\theta$ are taken as guiding rules $\rm{R}_{q} $, where N and $\theta$ are hyper-parameters.
Then, we append $q$ with one rule $r \in \rm{R}_{q}$ once at a time to avoid conflict and conduct rule-guided retrieval in the corpus $\mathcal{D}$ to obtain the top-k documents $\mathcal{D}^{r}_{q}$.
Finally, $\mathcal{D}^{r}_{q}$ from all rules in $\rm{R}_{q}$ are assembled to produce the final retrieval results $\mathcal{D}_{q}$, and RuleRAG-ICL conditions on the query $q$, rules $\rm{R}_{q}$ and documents $\mathcal{D}_{q}$ to reason the answer $a$.\looseness=-1

\textbf{Rule-guided retriever (RG-retriever).}  
The retriever calculates a relevant score $s(d_{i},q \circ r )$ between ($q$,$r$) and every document $d_{i} \in \mathcal{D}$:
$s(d_{i}, q \circ r )=\mathbf{E}_{d}(d_{i}) \cdot  \mathbf{E}_{q}\left (q \circ r \right ) $, where $\circ$ is sequence concatenation, $\cdot$ is dot product, $\mathbf{E}_{d}$ is document encoder and $\mathbf{E}_{q}$ is query encoder.
We select the top-k documents, $\mathcal{D}_{q}^{r}$,
and combine all $\mathcal{D}_{q}^{r}$ as the final retrieval results $\mathcal{D}_{q}$. This process is formalized as: 
\begin{equation}
\begin{aligned}
 & \mathcal{D}_{q} = \textstyle \bigcup\limits_{r\in\rm{R}_{q}} \mathcal{D}_{q}^{r};
& \mathcal{D}_{q}^{r}=
\text {arg} \ \ {\underset
{d_{i} \in \mathcal{D}}{\text {top}\mbox{-}\text {k}}} 
\ \ s\left(d_{i}, q \circ r \right ).
\end{aligned}
\end{equation}

\textbf{Rule-guided generator (RG-generator).} 
After recalling $\mathcal{D}_{q}$,
we construct an instruction to prompt LLMs to reason answers.
Different from implicit case-based prompts~\citep{wei2024instructraginstructingretrievalaugmentedgeneration}, we
directly inform LLMs of $\rm{R}_{q}$ as the attribution mechanisms and reason with the guidance of $\rm{R}_{q}$.
The probability of outputting answer $a$ can be approximated as:\looseness=-1
\begin{equation}
P (a \mid q, \mathcal{R},\mathcal{D}) \approx  P_{LLM}(a \mid \textsc{ins}(q, \rm{R}_{q}, \mathcal{D}_{q})),
\end{equation}
where $P_{LLM}\textsc{()}$ is the LLM generation probability. $\textsc{ins()}$ is instruction prompt, whose simplified form is in Figure~\ref{mainfigure} (c) and 
details are in Appendix~\ref{Prompt}.\looseness=-1

\subsection{RuleRAG-FT}
The overview of our proposed rule-guided retriever and generator fine-tuning is shown in Figure~\ref{mainfigure} (b) and (c).
For \textit{rule-guided retriever fine-tuning} (RGFT-retriever), we update the LM encoders in a contrastive learning objective~\citep{contrastive-learning} and train over supervised fine-tuning data 
provided in our constructed benchmarks
, where inputs are the queries plus rules and 
supervised labels
are heuristic oracle documents. 
For \textit{rule-guided generator fine-tuning} (RGFT-generator), we adopt the supervised instruction-tuning objective~\citep{iyer2023optiml} while combining query $q$ with two components: retrieved documents $\mathcal{D}_{q}$ and the set of rules $\rm{R}_{q}$ consistent with the retrieval phase. 
The rules introduced in the RGFT-generator train LLMs to optimally reason from the retrieved context into answers via attributable rules, making RuleRAG leverage our fine-tuned retrievers more rationally.

\textbf{Rule-guided retriever fine-tuning (RGFT-retriever).} We utilize two main types of retrievers: sparse and dense retrievers. As the sparse retriever, we use Pyserini to implement the standard training-free BM25~\citep{BM25}, which relies on word-level frequencies.
As the dense retrievers, we adopt the dual-encoder based retriever architecture, such as DPR and SimCSE.
We freeze the document encoder and tune the query encoder for high retrieval efficiency~\citep{NEURIPS2020_6b493230}. Given a $(\left (q , r \right ),\mathcal{D}_{o})$ pair in the fine-tuning data,
where $\mathcal{D}_{o}$ serve as the oracle documents, each $d_{i}^{+} \in \mathcal{D}_{o}$ is a positive learning example while each in-batch $d_{j}^{-} \not\in \mathcal{D}_{o}$ is a negative example. We
train the retrievers in an in-batch contrastive training fashion
with
the following loss function $\mathcal{L}^{r}_{q}$:
\begin{equation}
\mathcal{L}^{r}_{q}=
-\log \frac{e^{s({d}_{i}^{+}, q \circ r )}}{e^{s({d}_{i}^{+}, q \circ r )}+\sum_{d_{j}^{-} \epsilon \mathcal{B} / \mathcal{D}_{o}} e^{s({d}_{j}^{-}, q \circ r ) }},
\vspace{-0.02cm}
\end{equation}
where $\mathcal{B}$ is the documents for all the queries in one training batch. $\mathcal{D}_{o}$ is oracle documents for the query and $\mathcal{B} / \mathcal{D}_{o}$ is its in-batch negative examples. 
The final training goal of RGFT-retriever is to minimize the overall loss $\mathcal{L}= {\textstyle \sum_{ (\left (q , r \right ),\mathcal{D}_{o})\in\mathcal{F}_{R}}} \mathcal{L}^{r}_{q}$.

\textbf{Rule-guided generator fine-tuning (RGFT-generator).}  
To obtain greater model efficiency,
we fine-tune the generators in RuleRAG-FT, enhancing the proficiency to reason accurate answers following rules.
Formally, the designed instruction contains three parts: the relevant facts $\mathcal{D}_{q}$ retrieved by retrievers fine-tuned above, the rules $\rm{R}_{q}$ guiding 
attributable retrieval logics 
and the original query $q$. 
\looseness=-1

In practice, for open-source LLMs, we utilize the few-shot instruction fine-tuning strategy considering the following two aspects. 
First, our introduced rules reform the data-centric training to the alignment of task-centric abilities, i.e., it can be viewed as a reasoning task based on the guidance of rules~\citep{NEURIPS2023_ac662d74} 
and our training aim is to learn to use them.
Secondly, tuning all the data 
is prohibitive. We randomly select a fixed number of samples 
to conduct few-shot tuning 
(2048 samples in our practice). 
For closed-source LLMs, we perform 3-shot prompts as an empirical substitute of fine-tuning~\citep{dai2023why} due to the unavailable parameters. Specifically, we randomly select three $((q, \mathcal{D}_{q}, \rm{R}_{q}),$ $a)$ pairs 
as fixed examples in the prompts, making up the in-context augmentation.\looseness=-1



 \begin{table}[t]

 \belowrulesep=0pt
\aboverulesep=0pt
 \Huge
\centering
\resizebox{\linewidth}{!}{
\begin{tabular}{ccccccccc}
\toprule[0.10em]

&Benchmarks & $\left| \mathcal{R} \right|$ & $\left| \mathcal{D} \right|$ & $\left| \mathcal{F}_{R} \right|$ & $\left| \mathcal{F}_{G} \right|$ & $\left| \rm{Q} \right|$ &Source KG\\
\specialrule{0.05em}{1pt}{1pt}
\multirow{3}*{\rotatebox{90}{Temporal}}&RuleQA-I &557 &77,508 & 6,594&7,440&1,559&ICEWS14\\
&RuleQA-Y & 99& 243,633& 28,153&22,765&1,864& YAGO \\
&RuleQA-W &78 &584,364 & 50,996& 62,375&2,065&WIKI\\
 \specialrule{0em}{5pt}{5pt}
\specialrule{0.05em}{1pt}{1pt}
\multirow{2}*{\rotatebox{90}{Static}}&RuleQA-F & 367&49,088 &8,082 &9,645&1,233&FB15K-237 \\
&RuleQA-N & 234&18,177 &4,351 &4,764&815&NELL-995 \\
 \specialrule{0em}{1pt}{1pt}

 
\bottomrule[0.10em]
\end{tabular}}
 \caption{The statistics of our constructed benchmarks RuleQA. $\left| \mathcal{R} \right|$, $\left| \mathcal{D} \right|$, $\left| \mathcal{F}_{R} \right|$, $\left| \mathcal{F}_{G} \right|$ and $\left| \rm{Q} \right|$ are the numbers of rules, documents in corpus, retriever fine-tuning pairs, generator fine-tuning pairs and test queries, respectively.\looseness=-1}
\label{statistics}
\vspace{-0.4cm}
\end{table}

\section{Experimental Settings} \label{Experimental Setup}

\subsection{Benchmarks and Setup of RuleRAG}
\label{Setup}
The construction process of our constructed five rule-aware benchmarks RuleQA are in Appendix~\ref{Benchmarks}. The statistics of RuleQA are in Table~\ref{statistics}.
For RuleRAG-ICL, in addition to adding rule guidance to both retrievers and generators (RG-retriever + RG-generator), we also add rule guidance only to the retrieval stage (RG-retriever + generator), trying to prove that introducing rules in two stages can both contribute to the performance.
For RuleRAG-FT, the complete method involves retrievers and generators with RGFT.
The ablation study shows both of them are individually beneficial to the results.
To emphasize the contribution of rules, we introduce several variants of RuleRAG-FT. The SSFT in Table~\ref{Mainresults} represents the standard supervised fine-tuning following the vanilla manner, where 
the fine-tuning instruction consists only of the queries and retrieved documents without rules. 
Whether the inputs are added with rules during inference is consistent with how the models are fine-tuned.
\looseness=-1

 \begin{table*}[!]

 \belowrulesep=0pt
\aboverulesep=0pt
\Huge
\centering
\renewcommand\arraystretch{1.2}
\resizebox{\linewidth}{!}{
\begin{tabular}{ccc|ccc|ccc|ccc|ccc|ccc}

\toprule[0.10em]
&\multicolumn{2}{c}{Architecture }& \multicolumn{3}{c}{RuleQA-I }& \multicolumn{3}{c}{RuleQA-Y }& \multicolumn{3}{c}{RuleQA-W }& \multicolumn{3}{c}{RuleQA-F }& \multicolumn{3}{c}{RuleQA-N }\\
\cmidrule (lr){2-3}\cmidrule (lr){4-6}\cmidrule (lr){7-9}\cmidrule (lr){10-12}\cmidrule (lr){13-15}\cmidrule(lr) {16-18}
&Retriever &Generator &R@10&EM&T-F1&R@10&EM&T-F1&R@10&EM&T-F1&R@10&EM&T-F1&R@10&EM&T-F1\\
\specialrule{0.05em}{1pt}{1pt}
Standard Prompting &None &LLAMA2\_7B &- &1.5&19.4&- &0.4&12.4&- &1.5&27.7&- &1.0&24.9&- &0.1&10.4\\
Standard RAG &DPR &LLAMA2\_7B &14.1 &5.2&24.4&3.8 &2.6&18.5&7.4 &4.8&35.8&18.9 &11.0&33.1&19.3 &9.8&29.6\\
VE (3-shot) &DPR &LLAMA2\_7B &- &3.1&10.7&- &0.8&6.5&- &4.2&25.2&- &7.4&12.7&- &4.8&14.1\\
CoK (3-shot) &DPR &LLAMA2\_7B &- &4.0&12.5&- &1.9&10.4&- &5.7&29.0&- &9.8&18.7&- &7.4&21.6\\
\specialrule{0.05em}{1pt}{1pt}
\specialrule{0.05em}{1pt}{1pt}
\multirow{2}{*}{RuleRAG-ICL} &RG-DPR &LLAMA2\_7B &24.2 &5.5&25.1&6.6 &4.3&19.2&22.6 &10.9&37.1&29.9 &13.1&33.1&26.5 &11.1&30.6\\
&RG-DPR &RG-LLAMA2\_7B &\textbf{24.2} &\textbf{9.8}&\textbf{29.1}&\textbf{6.6} &\textbf{6.1}&\textbf{20.9}&\textbf{22.6} &\textbf{12.7}&\textbf{39.1}&\textbf{29.9 }&\textbf{19.0}&\textbf{35.7}&\textbf{26.5} &\textbf{15.2}&\textbf{32.8}\\
\specialrule{0.05em}{1pt}{1pt}
\specialrule{0.05em}{1pt}{1pt}
 RuleRAG-FT &RGFT-DPR &RGFT-LLAMA2\_7B &\textbf{45.1} &\textbf{20.5}&\textbf{38.9}&\textbf{55.7} &\textbf{44.6}&\textbf{41.6}&\textbf{49.9} &\textbf{41.6}&\textbf{47.5}&\textbf{95.1} &\textbf{34.9}&\textbf{48.4}&\textbf{92.5} &\textbf{42.0}&\textbf{57.9}\\
\specialrule{0.05em}{1pt}{1pt}
\multirow{18}{*}{}{\textit{Rule Ablation}} \\
\cmidrule (lr){2-18}
\multirow{3}{*}{variants of RuleRAG-FT} &SSFT-DPR &RGFT-LLAMA2\_7B &38.4 &18.7&38.4&46.5&41.5&38.4&39.3 &36.9&42.4&79.0 &31.5&47.3&80.7 &42.0&55.2\\
  &RGFT-DPR &SSFT-LLAMA2\_7B &45.1 &15.3&27.5&55.7 &43.7&33.2&49.9 &29.4&34.1&95.1 &14.2&29.6&92.5 &29.8&42.4\\
  &SSFT-DPR &SSFT-LLAMA2\_7B &38.4 &13.8&27.3&46.5 &37.4&33.8&39.3 &28.8&34.3&79.0 &12.0&27.1&80.7 &27.5&41.9\\
\specialrule{0.05em}{1pt}{1pt}
\multirow{18}{*}{}{\textit{RGFT Ablation}} \\
\cmidrule (lr){2-18}
\multirow{2}{*}{variants of RuleRAG-FT} &RG-DPR &RGFT-LLAMA2\_7B &24.2 &13.3&37.7&6.6 &13.9&25.6&22.6 &14.7&30.5&29.9 &21.6&36.7&26.5 &15.4&34.9\\
  &RGFT-DPR &RG-LLAMA2\_7B &45.1 &14.2&33.1&55.7&33.9&36.5&49.9 &38.7&43.4&95.1 &33.5&41.9&92.5 &37.2&47.6\\
\specialrule{0.05em}{1pt}{1pt}
\specialrule{0.05em}{1pt}{1pt}
RuleRAG-CoK &RG-DPR &RG-LLAMA2\_7B &- &5.1&17.9&- &2.6&14.5&- &8.4&32.2&- &11.8&26.1&- &9.2&25.7\\
 
\bottomrule[0.10em]

\end{tabular}}
 \caption{Performance comparison of RuleRAG-ICL, RuleRAG-FT and the variant of RuleRAG, RuleRAG-CoK. RG-DPR and RG-LLAMA2\_7B represent rule-guided DPR and rule-guided LLAMA2\_7B in RuleRAG-ICL. RGFT represents rule-guided fine-tuning in RuleRAG-FT. SSFT represents standard supervised fine-tuning. Standard Prompting does not have a retrieval stage, VE and CoK involve multiple search objects, which change several times, so there is no R@10. \textbf{The best performance of RuleRAG-ICL and RuleRAG-FT are in bold.}\looseness=-1}
 \label{Mainresults}
\vspace{-0.4cm}
\end{table*}

\subsection{Baselines}
Given that LLMs have lots of world knowledge, we report the performance of directly using LLMs as answer reasoners without retrieval (Standard Prompting in Table~\ref{Mainresults}). Additionally, we compare RuleRAG with three baselines based on RAG.
We instantiate the widespread RAG framework using off-the-shelf LLMs and retrievers with queries as input, standing for the standard RAG methods (Standard RAG in Table~\ref{Mainresults}, ~\ref{Contriever} and ~\ref{MoreLLM}).
Chain-of-thought (CoT) methods, verify-and-edit (VE; ~\citet{VE}) and chain-of-knowledge (CoK; ~\citet{cok}) correct outputs independently and sequentially respectively by leveraging external knowledge sources. Following their implementation, we initialize the knowledge sources as our corpus $\mathcal{D}$ and use 3-shot CoT prompts. 
Moreover, since RuleRAG relies solely on rule guidance instead of other sophisticated techniques like reflection or interleave, we also focus on the performance comparison of RuleRAG with and without rules.

\subsection{Evaluation Metrics}
For the retrieval stage, the quality of retrieved documents is critical for downstream queries and is usually measured by Recall@k~\citep{karpukhin-etal-2020-denseDPR}, indicating whether the top-k blocks contain targeted information. For our task, we calculate Recall@k (\textbf{R@k},\%) by checking whether the correct answer to the given query is contained in the retrieved top-k documents. The higher R@k, the more potentially useful retrievers are for generators.
For the generation stage, the quality of answers is measured by Exact Match (\textbf{EM},\%) and Token F1 (\textbf{T-F1},\%), which are widely recognized in QA performance evaluation~\citep{zhu2021retrieving}. For EM, an answer is deemed correct if its normalized form corresponds to any acceptable answer in the provided ground truth lists.
T-F1 treats the answers and ground truths as bags of tokens and computes the average token-level overlap between them~\citep{li-etal-2023-graph}.
\looseness=-1

\begin{figure*}[t]
    \centering
    \includegraphics
    [width=1\linewidth]
    {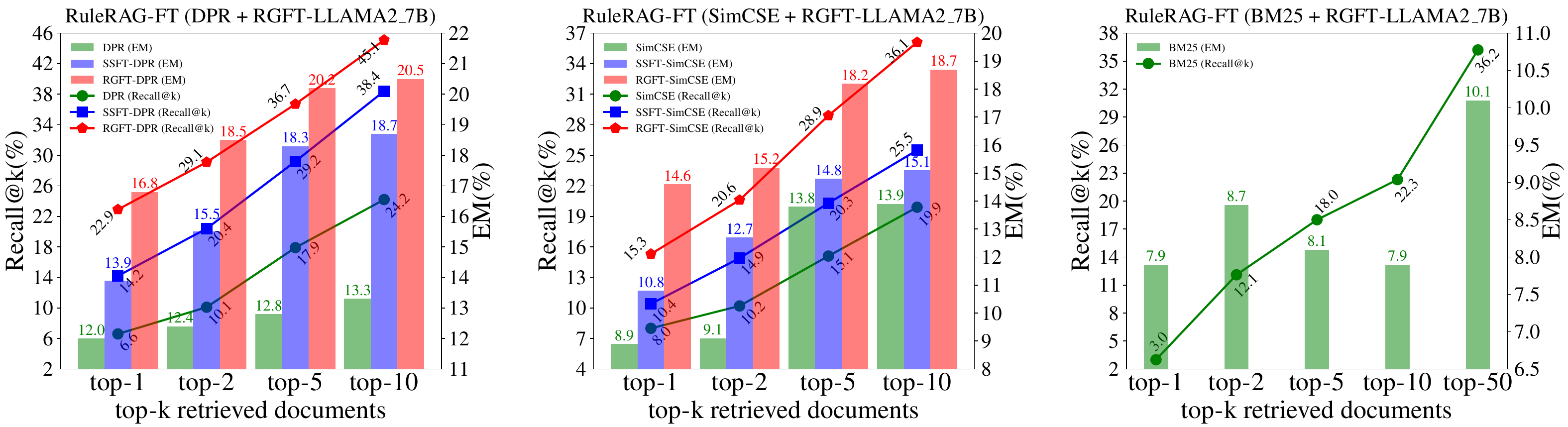}
     \vspace{-0.5cm}
    \caption{ The Reacll@k and EM performance of RuleRAG-FT in RuleQA-I with different numbers of retrieved documents and under multiple circumstances: three settings in DPR (DPR, SSFT-DPR and RGFT-DPR), three settings in SimCSE (SimCSE, SSFT-SimCSE and RGFT-SimCSE) and one setting in BM25. 
    Horizontal numbers over the pillars represent EM for bar charts and slanted numbers around the lines represent Recall@k for line charts.\looseness=-1
    }
    \vspace{-0.4cm}
    \label{moreretreiver}
\end{figure*}

\section{Experimental Results}

\subsection{Main Results}

Table~\ref{Mainresults} shows the overall experimental results in the five rule-aware QA benchmarks detailedly and provides a comprehensive comparison between our proposed RuleRAG-ICL, RuleRAG-FT, the variant of RuleRAG, RuleRAG-CoK, and all the baselines, under the instantiation of DPR~\citep{karpukhin-etal-2020-denseDPR} and LLAMA2\_7B~\citep{touvron2023llama} as retrievers and generators.
As a baseline without retrieval, LLAMA2\_7B using standard prompting can only refer to the knowledge it acquired during pre-training. Unsurprisingly, we notice that Standard Prompting (LLAMA2\_7B) yields the worst relative and absolute results in all the five benchmarks, revealing that parametric knowledge in LLMs makes it hard to answer our factual queries. Furthermore, the results of Standard Prompting avoid the concern that the performance improvement of subsequent experiments comes from intrinsic knowledge in LLMs. This also gives a side note to the challenges of our constructed five benchmarks and motivates the introduction of rules.\looseness=-1

The CoT-based methods, VE and CoK, use the rationales corrected by the retrieved knowledge to enhance the factual correctness of LLMs. From their results, it is evident that although they happen to succeed in modifying some answers by using rationales, they still fail to capture the logical relationships between the broader set of facts. 
The Standard RAG framework has better performance than the above non-retrieval or self-verifying methods, highlighting the importance of retrieved documents for knowledge-intensive queries.
However, their low performance is still unsatisfactory, suggesting that their principles of retrieval and generation are weak and leave much to be desired.
In the experiments, we illustrate that the performance can be further improved under the guidance of rules from two perspectives: through in-context learning (ICL) in RuleRAG-ICL and through RGFT in RuleRAG-FT.\looseness=-1

For RuleRAG-ICL (RG-DPR + LLAMA2\_7B), introducing rules in the retrieval stage alone enhances DPR recall performance and improves the answer accuracy of the original LLAMA2\_7B. RuleRAG-ICL (RG-DPR + RG-LLAMA2\_7B) consistently surpasses Standard RAG across various metrics (+9.3 in R@10, +5.9 in EM and +3.2 in T-F1 on average absolute performance over all five benchmarks),
achieving the improved performance. This confirms the sub-optimal ability of
the current RAG and the effectiveness of our proposed dual rule-guided retriever and generator. 
For RuleRAG-FT, our proposed RGFT can amazingly improve performance by a significant margin (+45.7 in R@10, +24.2 in EM and +15.3 in T-F1 compared to the best performance of RuleRAG-ICL).
In addition to Standard RAG-based RuleRAG-ICL and RuleRAG-FT, RuleRAG can also be applied to many advanced RAG-based models. As a variant of RuleRAG, RuleRAG-CoK introduces the idea of rule-guided RAG into CoK. The performance improvement achieved is attributed to our proposal. 

To further corroborate that these gains are due to the introduced rules, we first isolate the key component, rules, from fine-tuning data for RGFT, to form the standard supervised fine-tuning (SSFT) (\emph{Rule Ablation} in Table~\ref{Mainresults}) and then isolate the impact of the fine-tuned generator from the fine-tuned retriever in RuleRAG-FT (\emph{RGFT Ablation}  in Table~\ref{Mainresults}).
\emph{RGFT Ablation} shows both RGFT-DPR and RGFT-LLAMA2\_7B are beneficial when used individually, implicitly suggesting that the two phases do not depend on each other. Moreover, \emph{Rule Ablation} shows when we no longer leverage rules to explicitly inform the retrievers of the retrieval directions (SSFT-DPR) or how LLMs should correctly utilise the retrieved documents while fine-tuning (SSFT-LLAMA2\_7B), our recall and generation performances show varying degrees of degradation compared to RuleRAG-FT. This further clarifies the great assistance of rules
on the ability
to answer knowledge-intensive queries. 
\looseness=-1

\subsection{Further Analysis on Retrievers}
\indent\textbf{Retrievers in RuleRAG-FT}

In Figure~\ref{moreretreiver}, we initialize RuleRAG-FT with more retrievers: dense retrievers DPR~\citep{karpukhin-etal-2020-denseDPR}, SimCSE~\citep{gao-etal-2021-simcse} and training-free sparse retriever BM25~\citep{BM25}, and we use several retrieval configurations: retrievers without fine-tuning or with SSFT/RGFT while recalling different numbers of top-scored documents. Before fine-tuning, the Recall@k and EM performance of 
the three retrievers are comparable and each has their own performance characteristic, with no obvious advantages or disadvantages.
For instance, DPR has the best Recall@10 and SimCSE has the best EM under top-10 documents before fine-tuning.
\looseness=-1

After fine-tuning, DPR consistently outperforms SimCSE and RGFT consistently outperforms SSFT. Specifically, under considering top-scored documents with the same k, for the two trainable dense retrievers, the RGFT version recalls more relevant information (R@k) than the SSFT version by a large margin, demonstrating the generality of the proposed RGFT across different retrievers. 
As a result, the EM scores of the generated answers are better when higher-quality documents from retrievers are provided.
Moreover, when the retrievers and generators are applied with RGFT, RuleRAG-FT shows substantial performance gains, even with the retrieval number limited to top-1. For DPR and SimCSE, as we include more documents, the Recall@k and EM scores increasingly improve. This shows that leveraging rules to guide the retrieval and generation processes builds a bridge between queries and answers since rules provide retrieval directions and attributable mechanisms. For BM25, although Recall@k keeps increasing, EM experiences a drop, probably due to the retrieved noise.\looseness=-1

One additional finding is that even though the difference in R@2 between the original DPR and SimCSE is not large (10.1\% vs 10.2\%), the EM of generated answers can differ significantly (12.4\% vs 9.1\%). The reason may be that DPR's retrieved content includes not only the correct answers but also other helpful information. RGFT further widens the gap of Reacll@k between DPR and SimCSE.\looseness=-1

%

\noindent
\textbf{Retrievers in RuleRAG-ICL}

Contriever~\citep{Contriever} is a powerful retriever with strong unsupervised performance and can transfer well to new applications. Therefore, it has been widely used in RAG.
In Table~\ref{Contriever} in Appendix~\ref{ContrieverContriever}, we note that Contriever without the guidance of rules can achieve relatively good recall and RG-Contriever makes further enhancements. 
Compared to Standard RAG, RuleRAG-ICL with RG-Contriever and RG-generators also obtain varying degrees of performance improvement under the three LLMs.
These results confirm the outstanding ability of our proposed rule-guided method.

\subsection{Further Analysis on Generators}
\textbf{More LLMs as Generators}

To test RuleRAG's generalization to more generators,
we
evaluate the effect of different LLMs in Table~\ref{MoreLLM} in Appendix~\ref{Further}. 
We experiment with three more open-source LLMs: ChatGLM2\_6B~\citep{du2022glm}, Mistral\_7B\_v0.2~\citep{jiang2023mistral}, LLAMA2\_13B~\citep{touvron2023llama}, and a closed-source LLM, GPT-3.5-Turbo.
\looseness=-1

First, consistent with the conclusions for LLAMA2\_7B in Table~\ref{Mainresults}, Table~\ref{MoreLLM} show RuleRAG is effective under various kinds of LLMs. 
RuleRAG-ICL and RuleRAG-FT improve the overall performance of Standard RAG across all benchmarks and LLMs, demonstrating the validity and universality of rules. RuleRAG-FT consistently outperforms RuleRAG-ICL.
Secondly, for LLAMA2 as generators, Standard RAG, RuleRAG-ICL and RuleRAG-FT with the 13B model always outperform their 7B counterparts, indicating that the introduced rules can provide better guidance when using larger models with the same LLM architecture. Thirdly, the RuleRAG-ICL's EM results of GPT-3.5-Turbo are better than LLAMA2\_13B because of more massive model parameters, however, the RuleRAG-FT's EM results of LLAMA2\_13B are better than GPT-3.5-Turbo in three of the five benchmarks. This phenomenon illustrates that RGFT is fairly effective and necessary for lightweight LLMs to overcome big LLMs, making RuleRAG-FT much cheaper than off-the-shelf big LLMs for LLM deployment and application.\looseness=-1

\begin{figure}[t]
    \centering
    \includegraphics
    [width=0.75\linewidth]
    {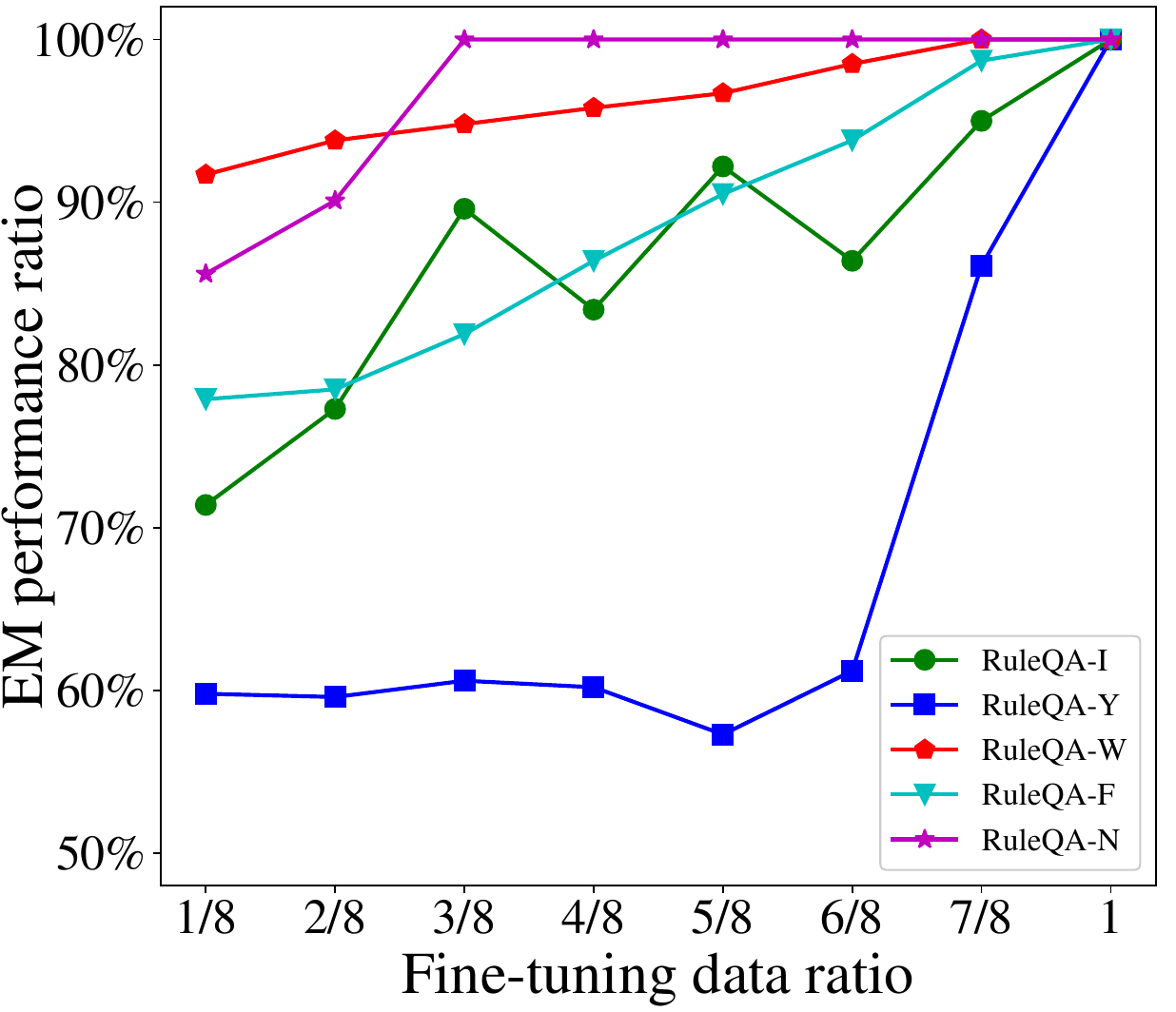}
    \caption{The EM variation of RuleRAG-FT produces different characteristics due to the varying difficulty of the rules in our constructed five RuleQA benchmarks. \looseness=-1
    }
    \vspace{-0.4cm}
    \label{trend}
\end{figure}

\noindent
\textbf{Impact of RGFT Data Volume}

In Figure~\ref{trend}, the x-axis is the ratio of fine-tuned data to the total amount of data in RGFT. The y-axis is the ratio of EM performance to the optimal one under DPR and LLAMA2\_7B, with closer to 100\% indicating stronger performance. 
Since the different properties of the rules in different benchmarks lead to different degrees of difficulty in learning, the growth of model performance under different benchmarks exhibits various characteristics.


The performance in RuleQA-Y fluctuates modestly at a very low level throughout the first half of the RGFT process,
and then sees a sudden surge in capability during the second half of the RGFT process.
It is worth noting that the EM performance in RuleQA-I fluctuates more dramatically: While realizing substantial EM performance gains (ranking second in all the benchmarks), it undergoes several upward and downward drops before levelling off at the optimal performance. This suggests that RuleQA-I is the most challenging among our constructed five benchmarks. Moreover, from Table~\ref{Mainresults}, ~\ref{Contriever} and ~\ref{MoreLLM}, we find RuleRAG has the worst absolute performance in RuleQA-I compared to the other four benchmarks under the same LLMs, which also illustrates the challenge of
the rules in 
RuleQA-I. \looseness=-1

\begin{figure}[t]
    \centering
    \includegraphics
    [width=0.78\linewidth]
    {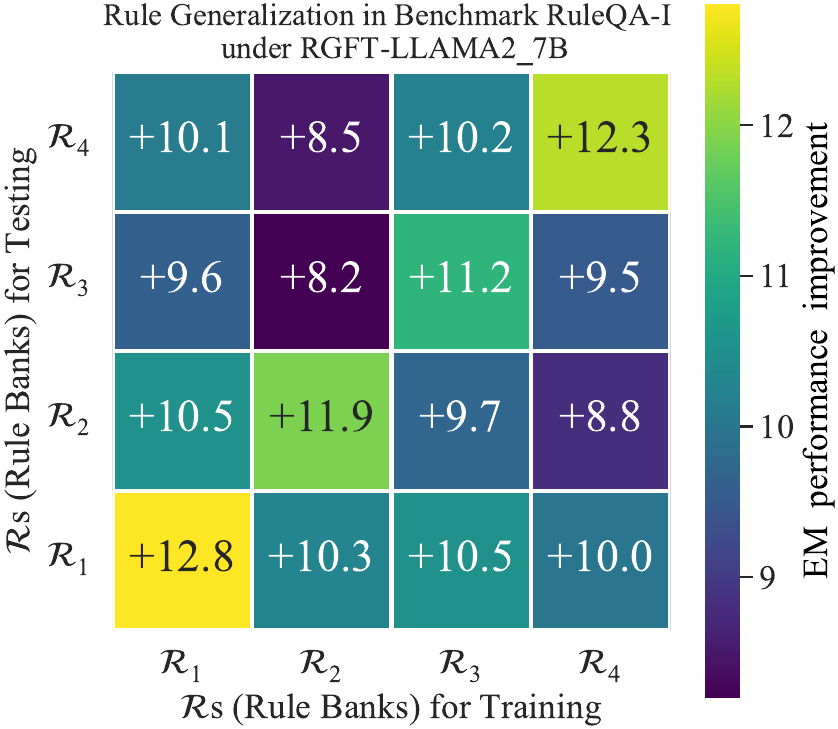}
    \caption{The EM of generalizing RuleRAG-FT from the source rule bank $\mathcal{R}_{i}$ to the target rule bank $\mathcal{R}_{j}$
        , i.e., 
        RuleRAG-FT is trained on $\mathcal{R}_{i}$ and tested on $\mathcal{R}_{j}$. 
        The numbers in ($\mathcal{R}_{i}$, $\mathcal{R}_{j}$) represent
        the performance gains compared to 
        the baseline Standard RAG tested on $\mathcal{R}_{j}$.
        \looseness=-1
    }
    \vspace{-0.6cm}
    \label{generalize}
\end{figure}

\section{Rule Generalization}
\textbf{Generalization on RuleQA}

RuleRAG-ICL is training-free, so we can attach arbitrary rules to the method's input by in-context learning. Experimental results above naturally illustrate its instruction-following ability to many kinds of rules.
In RGFT, the constructed fine-tuning data is limited anyway but rules are inexhaustible, so RuleRAG-FT cannot and should not see the full set of rules in RuleQA. Therefore, it is important to verify its ability to generalize to unseen rules.\looseness=-1


RuleRAG-FT must capture transferable rule utilization capability, since RuleRAG-FT has no prior knowledge of the target rule bank and is forced to learn from the source rule bank.
The results in Figure~\ref{generalize}, where $\mathcal{R}_{i}\cap\mathcal{R}_{j} = \emptyset$ and $\left|\mathcal{R}_{i}\right| = \left|\mathcal{R}_{j}\right| (i,j\in\{1,2,3,4\})$, show that 
(1) The diagonal ($\mathcal{R}_{i},\mathcal{R}_{i}$) has the highest performance gains and there are slight differences between various rule banks; (2) The results on two sides of the diagonal fluctuations within reasonable ranges and all show stable improvements over Standard RAG. 
This implies that RuleRAG-FT can
take advantage of the ability to leverage the learned underlying rule patterns
rather than being limited to concrete rule instances.
\looseness=-1




\noindent
\textbf{Generalization on More Datasets}

To test RuleRAG's performance on retrieving and reasoning using rules in a wider range of scenarios, we conduct assessments in four datasets: ASQA~\citep{stelmakh-etal-2022-asqa}, PopQA~\citep{mallen-etal-2023-trust}, HotpotQA~\citep{yang-etal-2018-hotpotqa} and Natural Questions (NQ)~\citep{kwiatkowski-etal-2019-natural}. 
Table~\ref{existing} shows RuleRAG's results on them. \looseness=-1

Even though these datasets were constructed without adapting rules, RuleRAG still achieves consistent performance gains with the help of the rules in our constructed RuleQA.
Specifically, following the framework of RuleRAG, existing datasets can be adaptively equipped with rules in RuleQA by calculating the relevance between candidate rules and queries. If some rules are highly relevant, they are introduced, otherwise no rules are introduced.
Table~\ref{existing} also compares the performance changes of an advanced RAG model CoK without and with our proposed  RuleRAG, indicating that when CoK replaces Standard RAG as the base method, the variant of RuleRAG, RuleRAG-CoK, still succeeds in introducing the guidance of rules.
These results further confirm the effectiveness of our proposed rule-guided retrieval and generation in RAG for more comprehensive QA models and applications.
\looseness=-1
 \begin{table}[t]
 \belowrulesep=0pt
\aboverulesep=0pt
\Huge
\centering
\resizebox{0.9\linewidth}{!}{
\begin{tabular}{ccccccc}
\toprule[0.10em]

Datasets  & ASQA & PopQA& HotpotQA& NQ   \\ 
\cmidrule(lr){2-5}
Methods  & EM  & EM&EM & EM  \\
\specialrule{0.05em}{1pt}{1pt}
\multicolumn{5}{c}{Contriever + LLAMA2\_7B}  \\
Standard RAG & 8.6  & 14.3&4.4   &7.6 \\
RuleRAG-ICL & 10.0  & 15.3 & 5.4& 7.8 \\
RuleRAG-FT & 11.1  &16.7 & 6.0 &8.1 \\
\specialrule{0.05em}{1pt}{1pt}
\multicolumn{5}{c}{Contriever + LLAMA2\_13B}   \\
Standard RAG &8.8   &13.7 & 5.8 &7.8 \\
RuleRAG-ICL& 10.4  &  17.3& 6.8& 8.3 \\
RuleRAG-FT& 11.9  & 18.8 &8.2 & 8.9\\
\specialrule{0.05em}{1pt}{1pt}
\multicolumn{5}{c}{Contriever + GPT-4o-mini}   \\
CoK&27.9&11.6&36.8&31.4\\
RuleRAG-CoK&40.0&16.2&38.6&35.4\\
\bottomrule[0.10em]
\end{tabular}}
 \caption{ The results of Standard RAG and CoK on four RAG datasets before and after equipping RuleRAG.\looseness=-1
 }
\label{existing}
\vspace{-0.6cm}
\end{table}



\section{Conclusion and Future Works}
In this paper, we point out two high-level problems of current RAG and propose rule-guided retrieval-augmented generation (RuleRAG). 
RuleRAG-ICL intuitively shows RAG can directly benefit from prompting LLMs with rules by in-context learning. To further improve the QA performance, RuleRAG-FT retrofits retrievers to recall more supportive information by contrastive learning and updates generators through our designed RGFT.
Experiments on our constructed five rule-aware QA benchmarks RuleQA show the strong performance of RuleRAG under multiple retrievers and generators and the generalization of rules. 
Furthermore, the comparison results with and without rules in RuleQA for RuleRAG and CoK on existing RAG datasets also attest to the effectiveness of rules in broader scenarios.
In the future, we will explore how to adapt rules in more complex RAG frameworks and 
use custom rules for more QA tasks.
\looseness=-1

\section*{Limitations}
Since existing RAG datasets do not have adapted rules, which have been widely used for knowledge-intensive reasoning tasks, we use mature KG rule mining algorithms to match rules for our constructed benchmarks RuleQA.
Although the experiments on four existing RAG datasets, including
ASQA, PopQA, HotpotQA and NQ,
initially demonstrated that the guideline of rules in RuleQA can be generalized to them and yielded performance gains, the gains were limited because the rules were not customized for them. Therefore, we plan to match rules for more RAG datasets and validate the rules on more RAG models to demonstrate the generic usefulness, since all RAG-based methods involve the two basic processes of retrieval and generation.




\bibliography{custom}
\newpage
\appendix

\section{Newly Constructed Rule-aware QA Benchmarks}
\label{Benchmarks}

Many real-world scenarios, such as healthcare, law and finance, rely on expert experience and the expertise can be represented with symbolic rules. For example, "having certain symptoms corresponds to a certain disease, which in turn requires the use of certain medications", "certain behaviours violate certain laws, which in turn require certain penalties", and "when approving a loan, information such as the borrower's debt-income ratio must be taken into account", and so on. In a broad sense, human common sense, expert experience, or regulations are rules and are ubiquitous in real life, but the common RAG datasets that are available today do not provide a corresponding rule base.
We find that knowledge bases and their rule mining algorithms can provide high-quality rules and question-answer pairs, so we construct RuleQA based on knowledge bases and equip rules to RuleQA.


\textbf{Rule bank $\mathcal{R}$.}
A huge amount of world knowledge, including static facts and temporal events, has been stored in static KGs and temporal KGs~\citep{jiang2023evolutionknowledgegraphssurvey}.
If \textit{ Event1} can lead to the happening of \textit{ Event2}, we believe that there is a logical correlation between them. In KGs, events are usually stored in the form of triples \textit{ [Entity 1, $r_{1}$, Entity 2] }, so we leverage a triple to stand for an event. 
In the static scenario, several different relations can be simultaneously established between two entities. In the temporal scenario, two entities can interact multiple times at different timestamps. Hence, if relation $r_{1}$ (rule body) can logically explain the occurrence of relation $r_{2}$ (rule head) between entities, we represent this relevance as rule $r$ in a natural language form: \textit{ [Entity 1, $r_{1}$, Entity 2] leads to [Entity 1, $r_{2}$, Entity 2]}.
We leverage the classical rule mining algorithm AMIE3~\citep{AMIE} for static KGs
and TLogic~\citep{TLogic} for temporal KGs. 
\emph{AMIE3 and TLogic are currently the most widely used rule mining methods, as well as one of the best static and temporal knowledge graph mining models, respectively. Therefore, the quality of the rules is guaranteed. They can intuitively give rules with confidence scores.
The frequently co-occur relations form rules with high confidence~\citep{liao-etal-2024-gentkg} and we only transform these high-confidence rules to the above text string form, comprising our rule bank $\mathcal{R}$, which will be consistently leveraged in the training and inferring process of RuleRAG.}
\looseness=-1

\textbf{Test dataset $\rm{Q}$.}
To avoid skewed entity distribution, we include links with both popular and long-tail entities in KG test sets and adjust their numbers to achieve balance. The remaining links are converted into queries with tail entities in these links as ground truths.
Different from PopQA~\citep{mallen-etal-2023-trust} 
with more low-popularity entities from Wikidata,
our benchmarks consider entities in uniform distribution from five knowledge bases, aiming to show the more general effectiveness of our method.
\looseness=-1

\textbf{Corpus $\mathcal{D}$ and fine-tuning datasets, $\mathcal{F}_{R}$ and $\mathcal{F}_{G}$.}
Different from EntityQuestions~\citep{sciavolino-etal-2021-simple}, we linearize the links in KG training sets into documents by concatenating entity, relation and time, forming concise and distinct factoids in $\mathcal{D}$, which serves as the retrieval source of RuleRAG. For RGFT, 
we split valid sets of KGs into two disjoint parts and convert the KG links of both parts into queries: one part is for queries in the fine-tuning datasets $\mathcal{F}_{R}$ for retrievers and the other part is for queries in the fine-tuning datasets $\mathcal{F}_{G}$ for generators.
Specifically, we search the corresponding oracle document examples from $\mathcal{D}$ for each
query-rule
pair
by entity name and relation-matching heuristics and take them as the golden training labels of the retrievers. 
Subsequently, we leverage the fine-tuned retrievers to retrieve relevant documents for each query in $\mathcal{F}_{G}$ and create fine-tuning instructions for generators by combining retrieval results, rules and queries, with golden answers as supervision. \looseness=-1

Benchmarks with temporal queries, named RuleQA-I, RuleQA-Y and RuleQA-W, are constructed based on three temporal KGs, ICEWS14~\citep{ICEWS14-0515}, YAGO~\citep{mahdisoltani2013yago} and WIKI~\citep{WIKI}. Benchmarks with static queries, named RuleQA-F and RuleQA-N, are constructed based on two static KGs, FB15K-237~\citep{FB15K-237} and NELL-995~\citep{NELL995}.


\section{Further Analysis on Retrievers}
\label{ContrieverContriever}
Table~\ref{Contriever} shows the performance of RuleRAG-ICL with Contriever and five LLMs.\looseness=-1

 \begin{table*}[t]
 \belowrulesep=0pt
\aboverulesep=0pt
\Huge
\centering
\renewcommand\arraystretch{1.2}
\resizebox{\linewidth}{!}{
\begin{tabular}{ccc|ccc|ccc|ccc|ccc|ccc}
\toprule[0.10em]
&\multicolumn{2}{c}{Architecture }& \multicolumn{3}{c}{RuleQA-I }& \multicolumn{3}{c}{RuleQA-Y }& \multicolumn{3}{c}{RuleQA-W }& \multicolumn{3}{c}{RuleQA-F }& \multicolumn{3}{c}{RuleQA-N }\\
\cmidrule (lr){2-3}\cmidrule (lr){4-6}\cmidrule (lr){7-9}\cmidrule (lr){10-12}\cmidrule (lr){13-15}\cmidrule(lr) {16-18}
&Retriever &Generator &R@10&EM&T-F1&R@10&EM&T-F1&R@10&EM&T-F1&R@10&EM&T-F1&R@10&EM&T-F1\\
\specialrule{0.05em}{1pt}{1pt}

Standard RAG &Contriever &LLAMA2\_7B &41.2 &18.7&36.2&52.7 &41.7&39.6&62.2 &45.5&51.2&80.6 &42.0&46.1&87.6 &45.2&56.5\\
\multirow{2}{*}{RuleRAG-ICL} &RG-Contriever &LLAMA2\_7B &45.5 &19.0&36.6&55.2 &42.6&42.3&63.2 &50.2&53.0&83.9 &43.9&50.0&88.5 &48.0&59.9\\
&RG-Contriever &RG-LLAMA2\_7B &45.5&22.8&39.6&55.2&47.8&43.0&63.2 &52.7&56.2&83.9&49.0&51.8&88.5&51.3&62.8\\


\specialrule{0.05em}{1pt}{1pt}
\specialrule{0.05em}{1pt}{1pt}

Standard RAG &Contriever &ChatGLM2\_6B &41.2 &8.5&24.7&52.7 &27.2&31.1&62.2 &41.6&42.9&80.6 &25.4&35.8&87.6 &4.9&8.3\\
\multirow{2}{*}{RuleRAG-ICL} &RG-Contriever &ChatGLM2\_6B &45.5 &10.5&25.4&55.2 &32.1&31.8&63.2 &43.8&43.2&83.9 &27.5&39.5&88.5 &12.0&12.8\\
&RG-Contriever &RG-ChatGLM2\_6B &45.5& 10.8&25.6&55.2& 32.9&32.5&63.2& 46.4&45.9&83.9&29.6&40.6&88.5& 16.5&14.2\\
\specialrule{0.05em}{1pt}{1pt}
\specialrule{0.05em}{1pt}{1pt}
Standard RAG &Contriever &Mistral\_7B\_v0.2 &41.2 &12.5&21.3&52.7 &37.8&36.1&62.2 &43.7&44.9&80.6 &21.5&36.8&87.6 &30.3&23.3\\
\multirow{2}{*}{RuleRAG-ICL} &RG-Contriever &Mistral\_7B\_v0.2 &45.5 &12.9&22.7&55.2 &38.4&37.5&63.2 &44.2&45.0&83.9 &23.9&38.2&88.5 &35.1&27.0\\
&RG-Contriever &RG-Mistral\_7B\_v0.2 &45.5 &15.4&24.5&55.2 &40.8&39.8&63.2 &46.3&45.8&83.9 &26.1&39.8&88.5 &39.8&31.9\\
\specialrule{0.05em}{1pt}{1pt}
\specialrule{0.05em}{1pt}{1pt}
Standard RAG &Contriever &LLAMA2\_13B &41.2 &22.1&39.5&52.7 &40.8&44.2&62.2 &49.2&52.4&80.6 &42.4&51.4&87.6 &50.2&57.4\\
\multirow{2}{*}{RuleRAG-ICL} &RG-Contriever &LLAMA2\_13B &45.5 &22.3&39.6&55.2 &41.1&44.4&63.2 &49.9&52.9&83.9 &45.0&52.0&88.5 &51.1&57.6\\
 &RG-Contriever &RG-LLAMA2\_13B &45.5 &22.3&39.8&55.2 &41.5&45.8&63.2 &51.2&54.2&83.9 &46.6&52.2&88.5 &52.7&58.1\\
 \specialrule{0.05em}{1pt}{1pt}
\specialrule{0.05em}{1pt}{1pt}
Standard RAG &Contriever &GPT-3.5-Turbo &41.2 &19.1&27.7&52.7 &38.1&44.2&62.2 &46.5&43.7&80.6 &56.3&39.1&87.6 &30.7&59.9\\
\multirow{2}{*}{RuleRAG-ICL} &RG-Contriever &GPT-3.5-Turbo &45.5 &19.7&30.1&55.2 &41.0&49.9&63.2 &49.4&65.8&83.9&56.5&50.3&88.5 &32.6&64.6\\
&RG-Contriever &RG-GPT-3.5-Turbo &45.5 &25.8&39.7&55.2&44.5&53.1&63.2 &53.1&68.7&83.9&57.6&59.0&88.5 &59.4&75.6\\
\bottomrule[0.10em]
\end{tabular}}
 \caption{The performance of RuleRAG-ICL with a powerful retriever, Contriever, under five LLMs.}

\label{Contriever}
\vspace{-0.3cm}
\end{table*}




\section{Further Analysis on LLMs}
\label{Further}
Table~\ref{MoreLLM} is the performance of RuleRAG-ICL and RuleRAG-FT with four more LLMs as generators.

 \begin{table*}[!]
 \belowrulesep=0pt
\aboverulesep=0pt
\Huge
\centering

\resizebox{\linewidth}{!}{
\begin{tabular}{ccc|cc|cc|cc|cc|cc}
\toprule[0.10em]
&\multicolumn{2}{c}{Architecture }& \multicolumn{2}{c}{RuleQA-I }& \multicolumn{2}{c}{RuleQA-Y }& \multicolumn{2}{c}{RuleQA-W }& \multicolumn{2}{c}{RuleQA-F }& \multicolumn{2}{c}{RuleQA-N }\\
\cmidrule (lr){2-3}\cmidrule (lr){4-5}\cmidrule (lr){6-7}\cmidrule (lr){8-9}\cmidrule (lr){10-11}\cmidrule(lr) {12-13}
&Retriever &Generator &EM&T-F1&EM&T-F1&EM&T-F1&EM&T-F1&EM&T-F1\\
\specialrule{0.05em}{1pt}{1pt}
Standard RAG &DPR &ChatGLM2\_6B &0.0 &5.1&0.3&7.8 &0.3&18.1&0.1 &21.0&0.0&0.0 \\
RuleRAG-ICL &RG-DPR &RG-ChatGLM2\_6B &2.5&16.9&1.3 &13.7&3.0 &26.7&10.8&27.3 &0.5&1.7\\
RuleRAG-FT &RGFT-DPR &RGFT-ChatGLM2\_6B &7.3 &21.2&42.2&35.2 &23.5 &30.5&19.2&29.8 &25.6&25.6\\
\specialrule{0.05em}{1pt}{1pt}
\specialrule{0.05em}{1pt}{1pt}
Standard RAG &DPR &Mistral\_7B\_v0.2 &1.6 &13.8&0.7&11.9 &1.3&21.8 &3.1&22.4&0.9 &1.5\\
RuleRAG-ICL &RG-DPR &RG-Mistral\_7B\_v0.2 &3.1&20.0&4.5 &23.4&34.2 &40.7&6.4&28.6 &4.2&16.6\\
RuleRAG-FT &RGFT-DPR &RGFT-Mistral\_7B\_v0.2 &22.6&34.9&49.2 &47.3&35.5&45.2&53.7&48.9 &50.9&62.6\\
\specialrule{0.05em}{1pt}{1pt}
\specialrule{0.05em}{1pt}{1pt}
Standard RAG &DPR &LLAMA2\_13B&6.1&25.9&4.0&20.2&6.0 &28.6&12.6&34.9 &10.2&31.6 \\
RuleRAG-ICL &RG-DPR &RG-LLAMA2\_13B &10.0&30.0&6.5 &23.7&14.1 &43.4&20.5&36.9 &18.2&36.1\\
RuleRAG-FT &RGFT-DPR &RGFT-LLAMA2\_13B &22.0&39.8&46.6 &47.9&42.3&48.1&45.6&49.6 &42.1&55.6\\
\specialrule{0.05em}{1pt}{1pt}
\specialrule{0.05em}{1pt}{1pt}
Standard RAG &DPR &GPT-3.5-Turbo &9.0&29.1&4.8 &25.9&6.9&31.5 &25.7&24.5 &16.0&43.3\\
RuleRAG-ICL &RG-DPR &RG-GPT-3.5-Turbo &12.2&30.3&9.9 &28.1&16.4 &33.7&37.9&32.1 &27.5&50.6\\
RuleRAG-FT &RGFT-DPR &RG-GPT-3.5-Turbo (3-shot)  &15.7&33.8 &40.1&32.8&38.9 &35.4&72.4&34.1 &68.1&56.1\\
\bottomrule[0.10em]
\end{tabular}}
\caption{The performance of RuleRAG-ICL and RuleRAG-FT with different LLMs as generators. The retriever is fixed as DPR. We omit R@10 since it has been given in detail in Table~\ref{Mainresults}. We use 3-shot prompts for the closed-source GPT-3.5-Turbo to replace RGFT due to its unpublished parameters.\looseness=-1}
\label{MoreLLM}
\vspace{-0.2cm}
\end{table*}

\section{Implementation Details}

\textbf{Generator fine-tuning.} We fine-tune the ChatGLM2\_6B, Mistral\_7B\_v0.2, LLAMA2\_7B, LLAMA2\_13B models using 2, 2, 4 and 8 V100 32G GPUs, respectively. We use LORA~\citep{hu2022lora} with 4-bit, a parameter-efficient fine-tuning (PEFT) adaptation method, to deal with the enormous computation costs and hardware requirements in training LLM. Hyper-parameter N is 3 and $\theta$ is 0.7. The fine-tuning hyperparameters are detailed in Table~\ref{hyperparameters}. Similar to \citet{lin2024radit}, we find that the best generalization performance on the dev set can be achieved using a small number of fine-tuning epochs. We evaluate the models every 3 epochs and select the best checkpoint based on the average dev set performance.

\textbf{Retriever fine-tuning.} We fine-tune DPR and SimCSE on 4 V100 32G GPUs using their public codes with a lr of 1e-5, a batch size of 32, and a temperature of 0.01. The base models are downloaded from their GitHub website.

 \begin{table*}[!]

 \belowrulesep=0pt
\aboverulesep=0pt
\centering
\resizebox{\linewidth}{!}{
\begin{tabular}{cccccccccc}
\toprule[0.10em]
LLM&lr&lora r &lora alpha& lora dropout&warm-up& batch size&epochs&model parallel&seq len\\
\cmidrule (lr){1-10}
ChatGLM2\_6B&3e-5&4&16&0.05&5&8&50&1&5120\\
Mistral\_7B\_v0.2&3e-5&4&16&0.05&5&8&50&1&5120\\
LLAMA2\_7B&3e-4&8&32&0.05&5&8&50&2&5120\\
LLAMA2\_13B&3e-4&16&32&0.05&10&4&50&4&5120\\
\bottomrule[0.10em]
\end{tabular}}
 \caption{Hyperparameters for RGFT-Generators.}
 \label{hyperparameters}
 \vspace{-0.5cm}
\end{table*}
\section{The Robustness of RuleRAG}
In the inferring process, since we can not know the content of the queries in advance, we may match some relevant rules for the queries regardless of whether the queries need the guidance of rules or not. In our preliminary experiments, we also find that, in some cases, retrieving information for some queries can directly match relevant documents. 

\textit{Therefore, in this section, we verify the robustness of our proposed method RuleRAG on queries which may not need the guidance of rules. We want to know if our introduced rules will interfere with the performance of retrieval and generation of such queries. }

Specifically, for each query in the benchmark, 
we degenerate it into a new relevant query by using the previously matched rules (\textit{ [Entity 1, $r_{1}$, Entity 2] leads to [Entity 1, $r_{2}$, Entity 2]}) and ensure that the answer is unchanged and that the relevant documents can be retrieved directly from the corpus. Meanwhile, according to the principle of performance comparison, we try to minimize interference with the original queries.
For instance, the original query is $\mathrm{What \ is \ the \ nationality \ of \ Jean\mbox{-}Luc \ Godard?}$ and the rule is that ``\textit{ [Entity 1, born in, Entity 2] leads to [Entity 1, has nationality, Entity 2]}''. Then, we convert the query into $\mathrm{Where \ is \ Jean\mbox{-}Luc \ Godard \ born?}$.
In this way, these queries can theoretically be successfully retrieved with related documents and correctly answered without the guidance of rules.\looseness=-1

In order to test the robustness of our rule-guided approach RuleRAG to such queries, we first conduct the Standard RAG on them as a baseline and then test the performance of RuleRAG by adding our previously matched rules. Hence, the only difference in the input of LMs between the main experiment and this experiment is the queries. The others, including rules and answers, remain the same.
The results are shown in Table~\ref{Robustness}.
We find (1) In terms of absolute performance, compared Table~\ref{Mainresults}, most of the results in Table~\ref{Robustness} show a certain degree of degradation, which indicates that \textit{we successfully achieve interference with the methods}. (2) Compared to the Standard RAG in Table~\ref{Robustness}, our proposed RuleRAG-ICL and RuleRAG-FT still achieve performance improvement over all the evaluation metrics, showing that our methods can overcome the interference of irrelevant rules. Fine-tuning based RuleRAG-FT is consistently better than RuleRAG-ICL, showing that our proposed RGFT is effective for these queries. Therefore, our methods are robust. 
\looseness=-1

To further improve the robustness of RuleRAG, in future work, we can use LLM to filter, sort, and evaluate rules or consider rules as interactable logical units, and so on. For exceptions or anomalies, we can also introduce entity linking for unrecognized entities and semantic similarity checks for outliers in temporal data. In addition, the robustness of the LLM itself can also ensure performance.

 \begin{table*}[!]
 \belowrulesep=0pt
\aboverulesep=0pt
\Huge
\centering

\resizebox{\linewidth}{!}{
\begin{tabular}{ccc|ccc|ccc|ccc|ccc|ccc}
\toprule[0.10em]
&\multicolumn{2}{c}{Architecture }& \multicolumn{3}{c}{RuleQA-I }& \multicolumn{3}{c}{RuleQA-Y }& \multicolumn{3}{c}{RuleQA-W }& \multicolumn{3}{c}{RuleQA-F }& \multicolumn{3}{c}{RuleQA-N }\\
\cmidrule (lr){2-3}\cmidrule (lr){4-6}\cmidrule (lr){7-9}\cmidrule (lr){10-12}\cmidrule (lr){13-15}\cmidrule(lr) {16-18}
&Retriever &Generator&R@10 &EM&T-F1&R@10&EM&T-F1&R@10&EM&T-F1&R@10&EM&T-F1&R@10&EM&T-F1\\
\specialrule{0.05em}{1pt}{1pt}

Standard RAG &DPR &LLAMA2\_7B&4.6 &10.7 &34.9 &2.7 & 3.6&19.7 &0.6 &2.3 &30.2 &15.2 &11.0 & 27.6&20.6 & 12.8&25.9  \\
RuleRAG-ICL &RG-DPR &RG-LLAMA2\_7B &11.8 &11.9 &35.5 & 5.3&9.5 &23.4 & 5.9&2.4 &32.5 &26.0 & 17.0& 39.9 &24.9 &17.6 &36.3 \\
RuleRAG-FT &RGFT-DPR &RGFT-LLAMA2\_7B &39.8 &16.6 &36.3  &46.8 &28.7 & 33.8&34.9 &15.9 &34.1 & 94.1&35.9 &48.9 &33.7 &20.4 &37.5 \\
\bottomrule[0.10em]
\end{tabular}}
\caption{The performance of RuleRAG-ICL and RuleRAG-FT for queries 
which may not need the guidance of rules to retrieve or generate. \textit{The results reflect the robustness of our methods.\looseness=-1}
}
\label{Robustness}
\vspace{-0.5cm}
\end{table*}

\section{The Choice of RuleRAG-ICL and RuleRAG-FT}
Our proposed RuleRAG includes two parts, RuleRAG-FT which requires training and RuleRAG-ICL which does not. They can also be used in combination with different LLMs: small-scale LLMs (6B, 7B, 13B in our paper) and a closed-source LLM (GPT-3.5-Turbo in our paper).

\textit{For different usage scenarios and requirements, we are free to choose different combinations. Summarizing all the results shown in this paper, we give the following heuristic decision criteria and corresponding reasons.}

Typically, the base performance of small-scale LLMs (the baseline Standard RAG) is low and the performance improvement of both RuleRAG-ICL and RuleRAG-FT with small-scale LLMs is very significant. Therefore, we can use the RuleRAG-ICL to get good results locally when hardware resources are limited. Otherwise, we recommend fine-tuning LLMs for better results. For our benchmarks, the inference time is 3-8 hours and the time for fine-tuning with the full data is 1-3 days. If users need to get inference results quickly in a short time, we recommend calling APIs of closed-source LLMs. In this combination, our methods' absolute performance and performance improvement are still very high (even optimal in some cases). For our benchmarks, their inference time is 0.5-2 hours.

\section{The EM performance Trend of LLAMA2\_7B and LLAMA2\_13B}

\begin{figure}[t]
    \centering
    \includegraphics
    [width=0.8\linewidth]
    {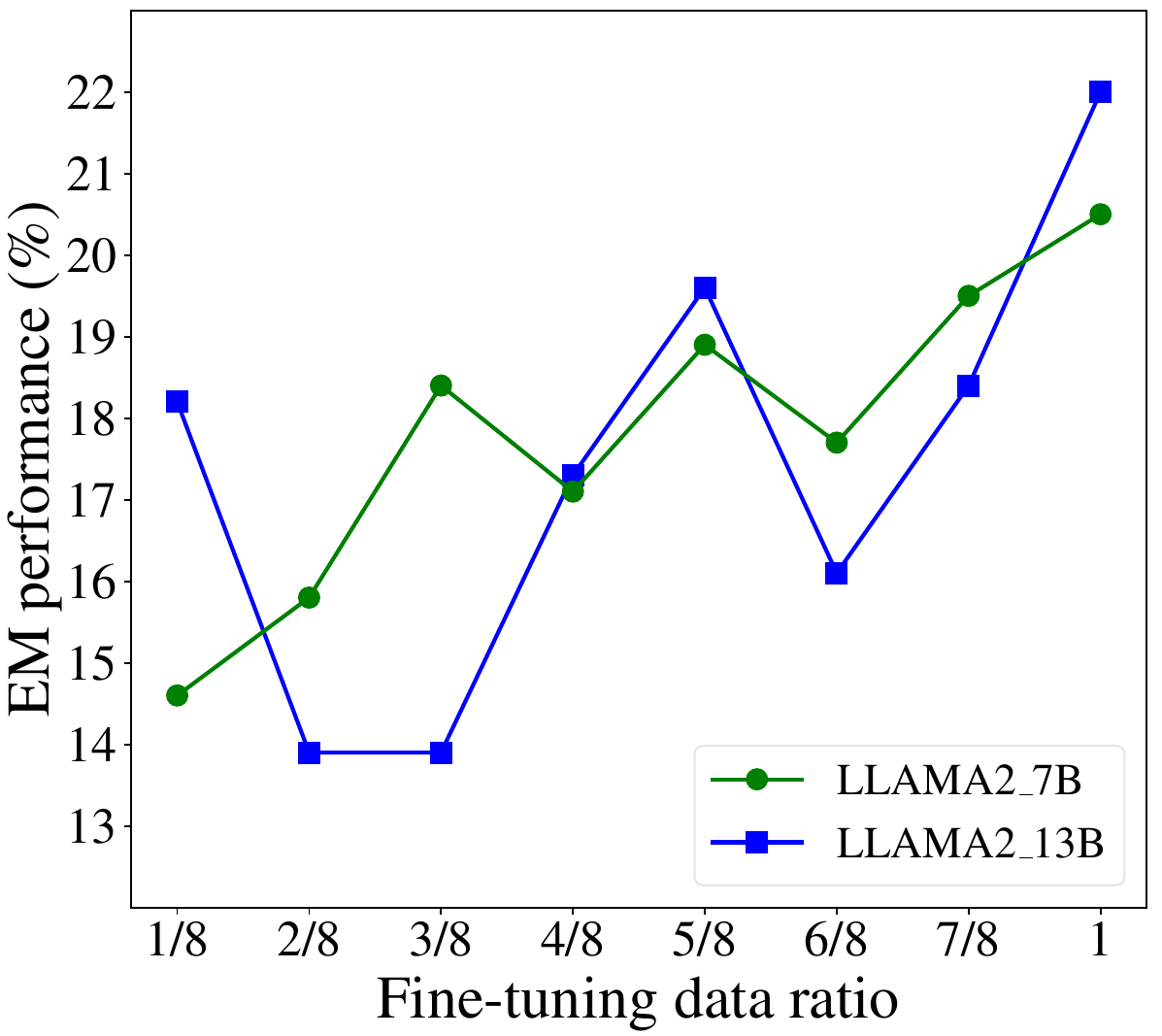}
    \caption{ The EM performance of RuleRAG-FT in RuleQA-I with RGFT-LLAMA2\_7B and RGFT-LLAMA2\_13B under increasing fine-tuning data ratio. The retriever is kept as RGFT-DPR.\looseness=-1
    }
    \vspace{-0.5cm}
    \label{2LLM}
\end{figure}

To make a stronger argument that dataset RuleQA-I is fairly difficult, we give in Figure~\ref{2LLM} how the EM performance of two different LLMs varies with the amount of fine-tuning dataset. From the figure, we find that the larger LLM ends up with better results (The result of LLAMA2\_13B is better than LLAMA2\_7B in the end), which is intuitive.
LLAMA2\_13B also experiences performance fluctuations, which illustrates the general challenging nature of RuleQA-I for multiple LLMs. In addition, we observe that in the second half of the fine-tuning process (the ratio from 4/8 to 1), both LLMs have similar change curves (up, then down, then up again), and the magnitude of change was greater for LLAMA2\_13B than for LLAMA2\_7B. We speculate that this is because both LLMs have similar model architectures, and thus the learning processes during fine-tuning are similarly guided; whereas, LLAMA2\_13B has more parameters, leading to fluctuating more and ultimately performing better.\looseness=-1

\begin{table*}[!htbp]

	\centering
	\begin{tabularx}{\textwidth}{X}
\bottomrule[0.2em]
\textbf{Query}:\\
Time 2014-12-11 what does Court Judge (Nigeria) Accuse ?\\
\specialrule{0.1em}{1pt}{1pt}
\textbf{Ground Truth}:\\
\textcolor{red}{Citizen (Nigeria).}\\
\specialrule{0.1em}{1pt}{1pt}
\textbf{Rules}:\\
Rule One: \\
\textcolor{white}{ }[Entity1, Accede to demands for change in leadership, Entity2] leads to [Entity1, Accuse, Entity2]. \\
Rule Two:\\ 
\textcolor{white}{ }[Entity1, Ease administrative sanctions, Entity2] leads to [Entity1, Accuse, Entity2].\\ 
Rule Three:\\ 
\textcolor{white}{ }[Entity1, Appeal for diplomatic cooperation, Entity2] leads to [Entity1, Accuse, Entity2].\\
\specialrule{0.1em}{1pt}{1pt}
\textbf{Retrieved documents of DPR (top-10)}:\\
\ding{172} Time 2014-08-22 Representatives (Nigeria) Consult Media (Africa). \ding{173} Time 2014-05-27 Activist (Nigeria) Consult Associated Press. \ding{174} Time 2014-06-16 Education (Nigeria) Consult Gabriel Torwua Suswam. \ding{175} Time 2014-09-03 Media (Nigeria) Consult Stephen Davis. \ding{176} Time 2014-05-21 Media (Nigeria) Consult Ministry (Nigeria). \ding{177} Time 2014-09-03 Media (Nigeria) Consult Stephen Davis. \ding{178} Time 2014-08-29 Media (Nigeria) Consult Stephen Davis. \ding{179} Time 2014-03-19 Citizen (Nigeria) Accuse Media (Nigeria). \ding{180} Time 2014-05-27 Activist (Nigeria) Consult Associated Press. \ding{181} Time 2014-10-07 Amnesty International Criticize or denounce Representatives (Nigeria).\\
\specialrule{0.1em}{1pt}{1pt}
\textbf{Retrieved documents of RG-DPR (top-10)}:\\
\ding{172} Time 2014-11-06 Court Judge (Nigeria) Arrest, detain, or charge with legal action Boko Haram. \ding{173} Time 2014-07-07 Court Judge (Nigeria) Make optimistic comment Nigerian Bar Association. \ding{174} Time 2014-10-01 Court Judge (Nigeria) Arrest, detain, or charge with legal action Boko Haram. \ding{175} Time 2014-06-12 Court Judge (Nigeria) Arrest, detain, or charge with legal action Citizen (Nigeria). \ding{176} Time 2014-07-21 Court Judge (Nigeria) Arrest, detain, or charge with legal action Citizen (Nigeria). \ding{177} Time 2014-04-11 Court Judge (Nigeria) Arrest, detain, or charge with legal action Citizen (Nigeria). \ding{178} Time 2014-08-26 Court Judge (Nigeria) Appeal for diplomatic cooperation (such as policy support) Citizen (Nigeria). \ding{179} Time 2014-04-04 Court Judge (Nigeria) Appeal for diplomatic cooperation (such as policy support) Government (Nigeria). \ding{180} Time 2014-09-16 Court Judge (Nigeria) Appeal for diplomatic cooperation (such as policy support) Citizen (Nigeria). \ding{181} Time 2014-07-08 Court Judge (Nigeria) Make optimistic comment Nigerian Bar Association.\\
\specialrule{0.1em}{1pt}{1pt}
\textbf{Retrieved documents of RGFT-DPR (top-10)}:\\
\ding{172} Time 2014-09-16 Court Judge (Nigeria) Appeal for diplomatic cooperation (such as policy support) Citizen (Nigeria). \ding{173} Time 2014-04-03 Court Judge (Nigeria) Appeal for diplomatic cooperation (such as policy support) Other Authorities / Officials (Nigeria). \ding{174} Time 2014-08-26 Court Judge (Nigeria) Appeal for diplomatic cooperation (such as policy support) Citizen (Nigeria). \ding{175} Time 2014-04-04 Court Judge (Nigeria) Appeal for diplomatic cooperation (such as policy support) Citizen (Nigeria). \ding{176} Time 2014-01-22 Court Judge (Nigeria) Ease administrative sanctions Citizen (Nigeria). \ding{177} Time 2014-09-16 Court Judge (Nigeria) Express intent to cooperate Citizen (Nigeria). \ding{178} Time 2014-07-17 Court Judge (Nigeria) Ease administrative sanctions Citizen (Nigeria). \ding{179} Time 2014-02-17 Court Judge (Nigeria) Ease administrative sanctions Member of Legislative (Govt) (Nigeria). \ding{180} Time 2014-02-28 Court Judge (Nigeria) Make an appeal or request Citizen (Nigeria). \ding{181} Time 2014-08-11 Court Judge (Nigeria) Make an appeal or request Citizen (Nigeria).\\
\specialrule{0.1em}{1pt}{1pt}
\textbf{Answer of Standard RAG (DPR + LLAMA2\_13B)}:\\
Media (Africa).\\
\specialrule{0.1em}{1pt}{1pt}
\textbf{Answer of RuleRAG-ICL (RG-DPR + RG-LLAMA2\_13B)}:\\
\textcolor{red}{Citizen (Nigeria).}\\
\specialrule{0.1em}{1pt}{1pt}
\textbf{Answer of RuleRAG-FT (RGFT-DPR + RGFT-LLAMA2\_13B)}:\\
\textcolor{red}{Citizen (Nigeria).}\\

\bottomrule[0.2em]
	\end{tabularx}%
     \caption{A detailed case study in RuleQA-I. We show the retrieved documents of three kinds of retrievers (DPR, RG-DPR, RGFT-DPR) and the answers of Standard RAG, RuleRAG-ICL and RuleRAG-FT with LLAMA2\_13B.}
      \label{casecase}
\end{table*}%

\section{The Difference of RuleQA and Existing RAG Datasets}

Most existing RAG datasets for QA only provide questions and corpora when construction and do not match suitable rules for them. In this paper, we construct five rule-aware QA benchmarks RuleQA, where many high-quality rules are mined from KGs to guide the retrieval and reasoning. Meanwhile, we experimentally show that both the introduced rules and our proposed model RuleRAG are effective in four existing RAG datasets, ASQA~\citep{stelmakh-etal-2022-asqa} (long-form QA), PopQA~\citep{mallen-etal-2023-trust}(short-form QA), HotpotQA~\citep{yang-etal-2018-hotpotqa}(multi-hop QA) and Natural Questions (NQ)~\citep{kwiatkowski-etal-2019-natural}.

In addition, although they are widely leveraged in evaluating the QA performance of LMs, we find that all these datasets are primarily focused on
multi-hop and comparison-type questions and pay less attention to queries that require logical thinking to reason. As we know, many queries in the real world are not justified by relevance alone, because in many cases the lexical level of relevance is not the information that can support the answer to the query, and even introduces a lot of noise instead. Therefore, in this paper, we construct five rule-aware QA benchmarks RuleQA based on five popular static KGs or temporal KGs to emphasize the importance of rules in the QA task. It is worth noting that our described construction way in Section~\ref{Benchmarks} is general and easy to reproduce. 
For newly defined rule patterns, we can quickly construct
corresponding benchmarks using the above construction way,
showing its better scalability.


Moreover, our constructed RuleQA also provide corresponding fine-tuning datasets, which aim to improve the retrieval and generation ability of LMs. Currently, obtaining high-quality and plentiful supervised data for a specific task is a challenging problem for researchers~\citep{wang2024learningplanretrievalaugmentedlarge}. 
Manual annotation is time-consuming and difficult to replicate. A very convenient and widely used way is to distil knowledge from LLMs. However, relying on LLMs to generate data for \emph{training} puts too much trust in them and does not actually guarantee the accuracy of the reasoning ability in the trained models. 

In contrast, in this paper, the fine-tuning datasets of the retrievers are obtained by pattern matching and retrieval recall; the fine-tuning datasets for generators are obtained by using the KG nodes as answers and using retrieved information as instructions. The entire process is efficiently streamlined and automatically generated.

\begin{table*}[t]

	\centering
	\begin{tabularx}{\textwidth}{X}
\bottomrule[0.2em]
\# Instruct: For the query in the form of ``Time \{time\} what does \{subject\} \{relation\} ?'', we provide a collection of text consisting of multiple documents in the form of ``Time \{time\} \{subject\} \{relation\} \{object\}.'' Your response should directly generate the missing \{object\}.\\
\# Retrieved documents: Documents related to the Query. Time 2014-06-23 Abdullah Abdullah Expel or withdraw peacekeepers Election Commission (Afghanistan). Time 2014-02-20 Abdullah Abdullah Make a visit Afghanistan. $\cdots$ Time 2014-07-16 Abdullah Abdullah Make a visit Ashraf Ghani Ahmadzai. $\cdots$ Time 2014-09-20 Abdullah Abdullah Make a visit Foreign Affairs (United States).\\
\# Rules: Use the following Two rules to answer the given Query.
Rule One: [Entity1, Abduct, hijack, or take hostage, Entity2] leads to [Entity1, Make a visit, Entity2]. Rule Two: [Entity1, Make a visit, Entity2] leads to [Entity1, Make a visit, Entity2].
\\
\# Query: Time 2014-12-01 what does Abdullah Abdullah Make a visit ?\\
\specialrule{0.1em}{1pt}{1pt}
\# Answer: Afghanistan.\\
\bottomrule[0.2em]
	\end{tabularx}%
     \caption{Instruct prompt.}
     
     \label{Instruct prompts}
\end{table*}%

\section{Case Study}
A concrete example in Table~\ref{casecase} visually compares the baseline model (Standard RAG) and our proposed methods, RuleRAG-ICL and RuleRAG-FT. 

Specifically, the documents retrieved by the original DPR are almost irrelevant to the query and only one out of the top 10 documents contains the correct answer ``Citizen (Nigeria)''. RG-DPR's retrieval results are more relevant to the query entity and semantically support the answer. Meanwhile, 5 of the top 10 documents contain the correct answer. The retrieval quality of the fine-tuned RGFT-DPR is the best. All the retrieved documents are strongly supportive while answering the query through the given rules. In addition, 8 out of the top 10 documents contain correct answers, which further reflects the strong performance of our proposed methods.

Moreover, in the answering stage, Standard RAG naturally obtains a wrong answer based on low-quality retrieval results. However, RuleRAG-ICL and RuleRAG-FT attribute the correct answer through in-context learning and fine-tuning under the guidance of the rules.

\section{Error Analysis}

We further analyzed the detailed performance of our proposed model on 60*5 incorrectly answered queries from the five benchmarks. There were three main classes of errors:

(a) Rule Failure (5\%): In the real world, rules can reflect the logical workings of most events. However, we cannot claim that absolutely no exceptions occur. Among the incorrect responses we sampled, we found that the answers to some questions did not follow the general rules of reasoning, which in turn resulted in response failures. Future work could address such special cases separately.

(b) Retrieval Error (55\%): In this section, we assume that a retrieval is considered correct as long as the correct answer is included in the top 10 recalled documents, and a retrieval is considered incorrect otherwise. 
Due to the very large size of the corpus and the large number of documents that are semantically similar but do not support the answer, even a fine-tuned retriever may not recall relevant facts for the correct answer. In almost all cases, the question can not be answered correctly if the retrieved documents are wrong.

(c) Attribution Error (40\%): Due to the complex logical relationships between events, when the retrieved documents contain the correct answer, the generator may still fail to follow the rules and then come up with an incorrect answer. Generally, the more documents in the top 10 retrieved information that are related to the correct answer, the higher the probability that the generator will answer correctly. The problem of attribution error occurs generally because there are only one to three supportive documents in the retrieved information.

\section{Prompt Templates}
\label{Prompt}
There are mainly two kinds of prompts in our model: prompts for fine-tuning in Figure~\ref{mainfigure} and prompts for in-context learning of GPT in Table~\ref{Contriever}. As Figure~\ref{mainfigure} shows, Instruct prompts consist of five parts: \emph{Instruct}, \emph{Retrieved documents}, \emph{Rules}, \emph{Query} and \emph{Answer}. The \emph{Instruct} is fixed, the \emph{Retrieved documents} are retrieved by our proposed RuleRAG according to \emph{Rules} and \emph{Query}, and the \emph{Answer} is pre-defined. As Section ~\ref{Experimental Setup} shows, we use 3-shot in-context learning for GPT to replace fine-tuning. In the following, we take RuleQA-I as an instance to show the RGFT instruct prompts (Table~\ref{Instruct prompts}) and prompts for GPT-3.5-Turbo (Table~\ref{GPT prompts}).

\begin{table*}[!htbp]
\small

	\centering
	\begin{tabularx}{\textwidth}{X}
 \bottomrule[0.2em]
Answer the Final Query by referring to the three cases below.\\\\
Case 1:\\
\# Instruct: For the query in the form of ``Time \{time\} what does \{subject\} \{relation\} ?'', we provide a collection of text consisting of multiple documents in the form of ``Time \{time\} \{subject\} \{relation\} \{object\}.'' Your response should directly generate the missing \{object\}.\\
\# Retrieved documents: Documents related to the Query. Time 2014-06-23 Abdullah Abdullah Expel or withdraw peacekeepers Election Commission (Afghanistan). Time 2014-02-20 Abdullah Abdullah Make a visit Afghanistan. $\cdots$ Time 2014-07-16 Abdullah Abdullah Make a visit Ashraf Ghani Ahmadzai. $\cdots$ Time 2014-09-20 Abdullah Abdullah Make a visit Foreign Affairs (United States).\\
\# Rules: Use the following Two rules to answer the given Query.
Rule One: [Entity1, Abduct, hijack, or take hostage, Entity2] leads to [Entity1, Make a visit, Entity2]. Rule Two: [Entity1, Make a visit, Entity2] leads to [Entity1, Make a visit, Entity2].\\
\# Query: Time 2014-12-01 what does Abdullah Abdullah Make a visit ?\\
\# Answer: Afghanistan.\\\\
Case 2:\\
\# Instruct: For the query in the form of ``Time \{time\} what does \{subject\} \{relation\} ?'', we provide a collection of text consisting of multiple documents in the form of ``Time \{time\} \{subject\} \{relation\} \{object\}.'' Your response should directly generate the missing \{object\}.\\
\# Retrieved documents: Documents related to the Query. Time 2014-04-07 Adams Oshiomhole Make an appeal or request Citizen (Benin). Time 2014-10-13 Adams Oshiomhole Accuse People's Democratic Party (Benin). $\cdots$ Time 2014-07-02 Adams Oshiomhole Criticize or denounce Citizen (Nigeria). $\cdots$ Time 2014-08-05 Adams Oshiomhole Praise or endorse Labor Union (Nigeria).\\
\# Rules: Use the following Three rules to answer the given Question. Rule One: [Entity1, Make an appeal or request, Entity2] leads to [Entity1, Make an appeal or request, Entity2]. Rule Two: [Entity1, Appeal for economic aid, Entity2] leads to [Entity1, Make an appeal or request, Entity2]. Rule Three: [Entity1, Accuse of aggression , Entity2] leads to [Entity1, Make an appeal or request, Entity2].\\
\# Query: Time 2014-12-01 what does Adams Oshiomhole Make an appeal or request ?\\
\# Answer: Citizen (Nigeria).\\\\
Case 3:\\
\# Instruct: For the query in the form of ``Time \{time\} what does \{subject\} \{relation\} ?'', we provide a collection of text consisting of multiple documents in the form of ``Time \{time\} \{subject\} \{relation\} \{object\}.'' Your response should directly generate the missing \{object\}.\\
\# Retrieved documents: Documents related to the Query. Time 2014-09-25 Adams Oshiomhole Demand Citizen (Benin). Time 2014-02-05 Adams Oshiomhole Express intent to cooperate Citizen (Nigeria). $\cdots$ Time 2014-10-13 Adams Oshiomhole Make an appeal or request Other Authorities / Officials (Nigeria). $\cdots$ Time 2014-07-01 Adams Oshiomhole Praise or endorse Media (Africa).\\
\# Rules: Use the following Three rules to answer the given Question. Rule One: [Entity1, Obstruct passage, block, Entity2] leads to [Entity1, Praise or endorse, Entity2]. Rule Two: [Entity1, Expel or deport individuals, Entity2] leads to [Entity1, Praise or endorse, Entity2]. Rule Three: [Entity1, Praise or endorse , Entity2] leads to [Entity1, Praise or endorse, Entity2].\\
\# Query: Time 2014-12-01 what does Adams Oshiomhole Praise or endorse ?\\
\# Answer: Media (Africa).\\\\
Final Query:\\
\# Instruct: For the query in the form of ``Time \{time\} what does \{subject\} \{relation\} ?'', we provide a collection of text consisting of multiple documents in the form of ``Time \{time\} \{subject\} \{relation\} \{object\}.'' Your response should directly generate the missing \{object\}.\\
\# Retrieved documents: Documents related to the Query. Time 2014-03-11 Alexis Tsipras Make a visit Ireland. Time 2014-02-26 Alexis Tsipras Express intent to meet or negotiate Slovenia. $\cdots$ Time 2014-05-26 Alexis Tsipras Make a visit Head of Government (Greece). $\cdots$ Time 2014-09-17 Alexis Tsipras Consult New Democracy.\\
\# Rules: Use the following Three rules to answer the given Question. Rule One: [Entity1, Accede to demands for change in leadership, Entity2] leads to [Entity1, Make statement, Entity2]. Rule Two: [Entity1, Demand release of persons or property, Entity2] leads to [Entity1, Make statement, Entity2]. Rule Three: [Entity1, Accuse of crime, corruption , Entity2] leads to [Entity1, Make statement, Entity2].\\
\# Query: Time 2014-12-01 what does Alexis Tsipras Make statement ?\\
\# Answer: \\
\bottomrule[0.2em]
	\end{tabularx}%
    \caption{GPT-3.5-Turbo prompt.}
\label{GPT prompts}
\end{table*}%

\end{document}